\documentclass[notitlepage,prd,twocolumn,showpacs,preprintnumbers,nofootinbib,superscriptaddress,floatfix,tightenlines]{revtex4-1}
\usepackage{mathrsfs}
\usepackage{amsmath,amssymb,amsfonts,amsbsy}
\usepackage{graphicx}
\usepackage{color}
\usepackage[paperwidth=210mm,paperheight=297mm,centering,hmargin=1.8cm,top=1.7cm,bottom=2.5cm]{geometry}
\usepackage{hyperref}

\newcommand{\invcoh}{Q}		% inverse coherence length
\newcommand{\pMq}{P}		% the integration variable $|\mathbf{p - q}|$
\newcommand{\pSqrMqSqr}{2m|\omega|}	% $|p^2 - q^2|$
\newcommand{\PiR}{\Pi_R}
\newcommand{\PiKernel}{\Gamma_{\Pi}}
\newcommand{\CKernel}{\Gamma_{C}}
\newcommand{\ttref}{t}
\newcommand{\trefTwo}{t_{\mathrm{ref}}}
\newcommand{\ggSim}{\gtrsim}
\newcommand{\llSim}{\lesssim}
\newcommand{\Pinv}{\pMq_{\mathrm{inv}}}
\newcommand{\FFunc}{I^{2\leftrightarrow 2}}
\newcommand{\fs}{f_S}
\newcommand{\Ks}{K_S}

\newcommand{\relsup}{^{\mathrm{rel}}}

\newcommand{\nrsup}{^{\mathrm{nr}}}

\begin{document}

\title{Large-$N$ kinetic theory for highly occupied systems}

\author{R.\ Walz}
\email{roland\_walz@t-online.de}
\affiliation{Institut f\"{u}r Theoretische Physik, Universit\"{a}t Heidelberg, Philosophenweg 16, 69120 Heidelberg, Germany}

\author{K.\ Boguslavski}
\email{kirill.boguslavski@jyu.fi}
\affiliation{Department of Physics, University of Jyv\"{a}skyl\"{a}, P.O.~Box 35, 40014 University of Jyv\"{a}skyl\"{a}, Finland}

\author{J.\ Berges}
\email{berges@thphys.uni-heidelberg.de}
\affiliation{Institut f\"{u}r Theoretische Physik, Universit\"{a}t Heidelberg, Philosophenweg 16, 69120 Heidelberg, Germany}

\begin{abstract}
We consider an effective kinetic description for quantum many-body systems, which is not based on a weak-coupling or diluteness expansion. Instead, it employs an expansion in the number of field components $N$ of the underlying scalar quantum field theory. Extending previous studies, we demonstrate that the large-$N$ kinetic theory at next-to-leading order is able to describe important aspects of highly occupied systems, which are beyond standard perturbative kinetic approaches. We analyze the underlying quasiparticle dynamics by computing the effective scattering matrix elements analytically and solve numerically the large-$N$ kinetic equation for a highly occupied system far from equilibrium. This allows us to compute the universal scaling form of the distribution function at an infrared nonthermal fixed point within a kinetic description and we compare to existing lattice field theory simulation results. 
\end{abstract}

\maketitle

%%%%%%%%%%%%%%%%%%%%%%%%%%%%%%%%%%%%%%%%%%%%%%%%%%%%%%%%%%%%%%%%%%%%%%%%%%%%%%%%%%%%%%%%%%%%%%%%%%%%
%%%%%%%%%%%%		SECTION: INTRODUCTION
%%%%%%%%%%%%%%%%%%%%%%%%%%%%%%%%%%%%%%%%%%%%%%%%%%%%%%%%%%%%%%%%%%%%%%%%%%%%%%%%%%%%%%%%%%%%%%%%%%%%

\section{Introduction}
\label{sec_introduction}

A fully microscopic description of the real-time dynamics of quantum many-body systems in terms of quantum field theory can be very demanding. Often, effective theories with a well-defined range of validity at some (long) time and distance scales provide an efficient alternative description. A well-known example is kinetic theory, which describes the state of the system in terms of a classical phase-space distribution of particles, $f(t,\mathbf{x},\mathbf{p})$, at time $t$ with position $\mathbf{x}$ and momentum $\mathbf{p}$~\cite{Pitaevskii:1981}. 

Accordingly, the derivation of kinetic theory from the underlying quantum field theory involves a series of crucial assumptions~\cite{Kadanoff:1962,Mueller:2002gd,Berges:2005md,Arnold:2007pg,Orioli:2015dxa}. An important condition is that the de Broglie wavelength $\sim 1/|\mathbf{p}|$ of relevant (quasi-)particles must be small compared to the mean free path between collisions. Otherwise, a description in terms of classical particles with a well-defined position and momentum between collisions would not be valid. Likewise, quantum interference effects between successive scattering events should not spoil a description in terms of independent scatterings. The existence of quasiparticle modes with a well-defined dispersion relation $\omega(\mathbf{p})$ translates in the language of quantum field theory to sufficiently narrow peaks of the spectral function~\cite{Blaizot:2001nr}. 

These conditions can be often met in the presence of a sufficiently weak coupling or small diluteness parameter controlling the strength of the scatterings. In particular, controlled perturbative kinetic descriptions exist for fermionic quantum field theories, and scalar field theories close to equilibrium where the relevant modes with momenta of the order of the temperature have occupancies of order one~\cite{Kadanoff:1962}. Likewise, perturbative descriptions exist far from equilibrium~\cite{Zakharov:1992,Micha:2004bv,Nazarenko:2011} if the occupancies of typical particle modes are not too high such that $f \ll 1/\lambda$ with $\lambda$ representing the relevant coupling constant or diluteness parameter. Though gauge theories are more involved, perturbative kinetic descriptions dealing with the problem of quantum interference have been given~\cite{Arnold:2002zm,Kurkela:2014tea,Kurkela:2015qoa}.

Much less is known about effective kinetic descriptions for general far-from-equilibrium situations. Pressing applications concern systems where the occupancies of relevant modes are non-perturbatively large ($f \sim 1/\lambda$) such that a perturbative power counting in terms of a small coupling parameter fails. 

An important example concerns the early stages of a relativistic heavy-ion collision (for recent reviews see \cite{Kurkela:2016vts,Fukushima:2016xgg}, and \cite{Baier:2000sb,Arnold:2002zm,Kurkela:2014tea,Kurkela:2015qoa} for current perturbative kinetic descriptions). In this situation, typical gauge boson occupancies can become non-perturbatively large at low momenta below the Debye mass scale. These modes may influence the evolution of important quantities like the longitudinal pressure $P_L$ of the expanding plasma, evidence of which was found from real-time lattice simulations~\cite{Berges:2015ixa,Berges:2013eia,Berges:2013fga}. Recent studies have found universal scaling behavior of infrared modes~\cite{Berges:2017igc} that may be connected to nontrivial field configurations~\cite{Mace:2016svc}. The possible existence and influence of an enhanced low-momentum region for non-Abelian plasmas out of equilibrium has been extensively discussed in the literature~\cite{Blaizot:2011xf,Kurkela:2012hp,Blaizot:2013lga,Xu:2014ega,York:2014wja,Blaizot:2016iir,Zhou:2017zql,Tanji:2017suk}, as well as methods for accessing spectral information at the Debye scale and below~\cite{Lappi:2016ato,Kurkela:2016mhu,Pawlowski:2016eck,Pawlowski:2017hpe}. 

Remarkably, longitudinally expanding non-Abelian plasmas and self-interacting scalar field theories are found to share important universal aspects of their far-from-equilibrium evolution~\cite{Berges:2014bba,Berges:2015ixa}. Similar self-similar scaling properties are known for a wide variety of highly occupied systems. These include relativistic scalar systems, often used in inflationary models of the early universe after a period of resonant particle production~\cite{Traschen:1990sw,Kofman:1994rk,Micha:2004bv,Berges:2008wm}, and non-relativistic systems such as ultracold quantum gases or other condensed matter systems after a strong quench~\cite{Berloff:2002,Scheppach:2009wu,Nowak:2011sk,Nowak:2013juc}. Characteristic infrared properties of these highly occupied systems turn out to be quantitatively the same for both relativistic and non-relativistic models~\cite{Schmiedmayer:2013xsa,Orioli:2015dxa}. The universal scaling properties are associated to a nonthermal renormalization group fixed point \cite{Berges:2008wm,Berges:2008sr} defining a universality class out of equilibrium, which encompasses non-relativistic and relativistic $N$-component field theories \citep{Orioli:2015dxa,Moore:2015adu}, scalar systems in different geometries \cite{Berges:2015ixa}, in different spatial dimensions \cite{Berges:2010ez,Karl:2016wko}, and it can even be observed for attractive quartic interactions as long as mean interactions are repulsive~\cite{Berges:2017ldx}. 
Perturbative kinetic approaches~\cite{Zeldovich:1969xx,Semikoz:1994zp,Semikoz:1995rd} break down at such large typical occupation numbers $f \gtrsim 1/\lambda$ and are not able to reproduce key features of this low-momentum dynamics~\citep{Orioli:2015dxa}. 

In order to describe the evolution also for non-perturbatively large occupation numbers, we consider an effective kinetic description for scalar systems which is not based on a weak-coupling or diluteness expansion. Developed in~\cite{Orioli:2015dxa,Berges:2010ez}, it exploits the fact that often one describes complex many-body problems with more than one particle species. In this case, alternative kinetic descriptions with an extended range of validity may be derived based on non-perturbative expansions in the number of species available. For scalar field theories with $N$ species and quartic self-interactions, it results from a large-$N$ expansion to next-to-leading order (NLO) based on a two-particle irreducible (2PI) resummation of self-energy diagrams~\cite{Berges:2001fi,Aarts:2002dj,Berges:2008wm,Berges:2008sr,Scheppach:2009wu}, which translates to a vertex resummation in the kinetic framework~\cite{Orioli:2015dxa,Berges:2010ez}. 

So far, the large-$N$ kinetic theory at NLO has been successfully applied to analytically compute the self-similarity exponents near nonthermal fixed points at low momenta, well agreeing with lattice results~\cite{Orioli:2015dxa}. However, a complete characterization of the non-perturbative infrared regime involves also the scaling form of the distribution function, which has not been established from the large-$N$ kinetic theory yet. In this work, we present the first numerical solution of the large-$N$ kinetic equation applied to the universal low-momentum scaling regime in three spatial dimensions. Our results are found to compare rather well to available lattice simulation data of the underlying field theory, in particular, establishing a $\sim |\mathbf{p}|^{-4}$ tail of the distribution in the regime dominated by number conservation.
We analyze in detail the range of validity and quasiparticle picture of the large-$N$ or ``vertex-resummed'' kinetic theory, and show how it encompasses and extends standard perturbative descriptions.  

The paper is organized as follows: In Sec.~\ref{sec:RelVsNonrel} we consider scalar $N$-component field theory. Starting from relativistic models with quartic self-interaction, we discuss the non-relativistic low-energy limit relevant, e.g., also for the description of ultracold Bose gases. Sec.~\ref{sec:pertkin} summarizes main aspects of perturbative kinetic theory, before we present the large-$N$ kinetic description in Sec.~\ref{sec:largeN}. The latter has an extended range of validity based on the inclusion of vertex corrections, which is analyzed in detail in Secs.~\ref{sec_scaling_effective_coupling} and \ref{sec_applicability}. We present a numerical solution of the large-$N$ kinetic theory for the description of a nonthermal fixed point in Sec.~\ref{sec_solution}. After concluding in Sec.~\ref{sec_conclusion}, we end with two appendices on calculational details of the collision integrals (App.~\ref{sec_preparation}) and on integration boundaries (App.~\ref{sec_integration_boundaries}).

\section{Relativistic and non-relativistic scalar fields}
\label{sec:RelVsNonrel}

We consider an $O(N)$ symmetric quantum field theory for the field components $\varphi_a(t,\mathbf{x})$, $a=1,\ldots,N$ with time $t$ and space variable $\mathbf{x}$ in three dimensions and quartic self-interactions. For $N=4$ the relativistic model describes the Higgs sector of the Standard Model of particle physics~\cite{Donoghue:1994}. In the context of low-energy descriptions of quantum chromodynamics, such a model encodes the three pions and the sigma resonance. Inflaton models for early universe cosmology often employ related multi-component field theories~\cite{Kofman:2008zz}. In a non-relativistic setting, the Heisenberg magnet for $N=3$ is a prominent example, and the case $N=2$ can be used to describe the two real components of a complex Bose field in systems of ultracold atomic gases dominated by $s$-wave scattering~\cite{Pitaevskii:2003}.

The considered relativistic quantum theory is described, on a classical level, by the action
\begin{equation}
S[\varphi] = \int_{t,\mathbf{x}} \left[ \frac{1}{2} \partial^\mu \varphi_a \partial_\mu \varphi_a - \frac{m^2}{2}  \varphi_a \varphi_a - \frac{\lambda}{4! N} \left( \varphi_a \varphi_a \right)^2\right]
\label{eq:classicalaction}
\end{equation}
with the notation $\int_{t,\mathbf{x}} \equiv \int \mathrm{d} t \int \mathrm{d}^3 x$ and
the (renormalized) mass $m$ and coupling parameter $\lambda$. Here summation over repeated Lorentz indices $\mu = 0,\ldots,3$ and field indices $a= 1,\ldots,N$ is implied. We will always employ natural units, with the speed of light, Boltzmann's constant and the reduced Planck constant equal to unity, i.e.\ $c=k_{\mathrm{B}}=\hbar =1$. 

For processes with characteristic momenta below the mass scale $m$, one may expect an effectively non-relativistic description to become relevant even for the relativistic microscopic model (\ref{eq:classicalaction}).  More generally, the descriptions of ultracold quantum gases or other condensed matter systems typically employ non-relativistic field theories. One may have in mind the phenomenologically important case of an $N = 2$ component non-relativistic field theory, which can equivalently be described in terms of a complex field $\phi(t,\mathbf{x})$. Following standard procedures~\cite{Berges:2015kfa}, the effective low-energy description may then be characterized by the non-relativistic action
\begin{equation}
S^{\mathrm{nr}}[\phi,  \phi^*] = \int_{t,\mathbf{x}} \left[ \phi^* \left( i \partial_t + \frac{\nabla^2}{2m} \right) \phi - \frac{g}{2} \left( \phi^* \phi \right)^2\right] ,
\label{eq:classicalaction_nr}
\end{equation}
Here we also introduced the effective non-relativistic coupling $g$, which is no longer dimensionless and may be related to the relativistic parameters as~\cite{Orioli:2015dxa}
\begin{equation}
g \sim \frac{\lambda}{m^2} \, .
\label{eq:glambda}
\end{equation}
For dilute Bose systems, $g$ can be related to the $s$-wave scattering length, given by $a=mg/(4\pi)$~\cite{Berges:2015kfa}. 

For the non-relativistic quantum theory in Eq.~\eqref{eq:classicalaction_nr} the expectation value of the particle density $n= \langle \phi^*\phi \rangle$ is conserved, and we will consider spatially homogeneous systems. The density $n$ and scattering length $a$ can be used to define a characteristic ``coherence length'' whose inverse is the momentum scale
\begin{align}
\label{eq_scale_M_def}
\invcoh = \sqrt{16 \pi\,a\,n} \sim \sqrt{m\,g\,n}\,.
\end{align}
We also define the ``diluteness parameter'' 
\begin{align}
\label{eq_diluteness_param_def}
\zeta = \sqrt{n\,a^3} \sim \invcoh\,m\,g\,,
\end{align}
which provides a dimensionless expansion parameter for the non-relativistic system, similar to the dimensionless coupling $\lambda$ for the relativistic system. Specifically, with (\ref{eq:glambda}) we obtain $\zeta \sim (Q/m)\, \lambda$. 

We emphasize that for the relativistic theory particle number is not conserved in general. However, for the highly occupied system considered in Sec.~\ref{sec_solution} an approximately conserved particle number is dynamically generated at a nonthermal fixed point such that the non-relativistic theory and the relativistic one can be in the same universality class of infrared scaling phenomena~\cite{Orioli:2015dxa}.

\section{Perturbative kinetic theory}
\label{sec:pertkin}

The derivation of perturbative kinetic equations from the underlying quantum-statistical field theory employs an expansion in terms of a small coupling $\lambda \ll 1$ or small diluteness parameter $\zeta \ll 1$, together with a gradient expansion for not too early times~\cite{Berges:2005md,Mueller:2002gd}. The phase-space distribution function of particles, $f(t,\mathbf{p})$, describing the state of the spatially homogeneous system at time $t$ and momentum $\mathbf{p}$, is obtained from the expectation value of two-field correlators evaluated at equal times.

More precisely, for the relativistic field theory the time derivative of the anti-commutator expectation value $\langle\{ \varphi_a, \varphi_b \}\rangle \equiv \langle \varphi_a \varphi_b + \varphi_b \varphi_a\rangle$ determines (in the absence of external forces) the change of the distribution function according to~\cite{Berges:2010ez}
\begin{equation}
\int_0^\infty \frac{\mathrm{d} \omega}{2\pi}\, \omega\, \frac{\partial}{\partial t} \left\langle \left\{ \varphi_a, \varphi_b \right\} \right\rangle (t,\omega, \mathbf{p}) \equiv \frac{\partial f(t,\mathbf{p})}{\partial t}\, \delta_{ab}\, . 
\label{eq:dfdt}
\end{equation} 
Here the frequency $\omega$ and spatial momentum $\mathbf{p}$ arise from the Fourier transform with respect to the relative space-time arguments of the two fields, while the remaining time dependence describes the breaking of time-translation invariance of the spatially homogeneous system out of equilibrium. The kinetic description involves the projection onto positive frequency contributions by integrating over $\omega$, respectively, and we exploit $O(N)$ symmetry assuming no spontaneous symmetry breaking such that $\langle\{ \varphi_a, \varphi_b \}\rangle \sim \delta_{ab}$.  

The kinetic equation may then be computed perturbatively by taking into account interaction effects, which are subsumed into the ``collision term'' $C[f]$ to obtain 
\begin{equation}
\frac{\partial f(t,\mathbf{p})}{\partial t} = C[f](t,\mathbf{p}) \, .
\label{eq_effective_Boltzmann_equation}
\end{equation}
In its range of validity, the leading contributions to $C[f]$ in a coupling expansion and an expansion to lowest order in gradients 
for the massive scalar field theory (\ref{eq:classicalaction}) leads to the well-known Boltzmann equation with 
the collision integral for elastic $2 \leftrightarrow 2$ scatterings~\cite{Berges:2015kfa},
\begin{align}
&C\relsup[f](t,\mathbf{p}) = \int_\mathbf{l,q,r} \frac{\lambda^2 (N+2)}{6 N^2} \, \FFunc[f](t,\mathbf{p,l,q,r}) \nonumber\\
& \times (2\pi)^4\delta^{(3)}(\mathbf{p+l-q-r})\frac{\delta(\omega\relsup_\mathbf{p} + \omega\relsup_\mathbf{l} - \omega\relsup_\mathbf{q} - \omega\relsup_\mathbf{r} )}{2 \omega\relsup_\mathbf{p}\,2 \omega\relsup_\mathbf{l}\,2 \omega\relsup_\mathbf{q}\,2 \omega\relsup_\mathbf{r}},
\label{eq_collision_integral_rel_perp}
\end{align}
with the notation $\int_\mathbf{q} \equiv \int \mathrm{d}^3 q/(2 \pi)^3$ and the relativistic dispersion
\begin{equation}
\omega\relsup_\mathbf{p} = \sqrt{m^2 + \mathbf{p}^2} \, .
\label{eq:reldispersion}
\end{equation} 
The functional $\FFunc[f]$ contains the distribution functions $f_\mathbf{p} \equiv f(t,\mathbf{p})$ describing the changes by loss or gain through scattering: 
\begin{align}
&\!\!\!\!\!\!\!\!\! \FFunc[f](t,\mathbf{p,l,q,r}) \nonumber \\
=\;\;\,& \left( f_\mathbf{p} + 1 \right) \left( f_\mathbf{l} + 1 \right)f_\mathbf{q} \, f_\mathbf{r} - f_\mathbf{p} \, f_\mathbf{l} \left( f_\mathbf{q} + 1\right) \left(f_\mathbf{r} + 1 \right) \nonumber \\
\overset{f \gg 1}{\approx}\;& \left( f_\mathbf{p} + f_\mathbf{l} \right) f_\mathbf{q} \, f_\mathbf{r} - f_\mathbf{p} \, f_\mathbf{l} \left( f_\mathbf{q} + f_\mathbf{r} \right)\,.
\label{eq_functional_F}
\end{align}
In the last equation, we give the approximate expression for large occupancies that will be useful later. 

Since the overall collision term $C[f]$ is of order $\lambda^2$, further perturbative corrections to terms appearing in the integrand of (\ref{eq_collision_integral_rel_perp}) are subleading. In particular, it allows one to employ in the integrand a well-defined dispersion relation (\ref{eq:reldispersion}). Phrased in terms of the underlying field theory, this leads to a quasiparticle form for the spectral function given by the expectation value of the commutator of two fields~\cite{Zakharov:1992,Berges:2010ez}:
\begin{equation}
\left\langle \left[ \varphi_a, \varphi_b \right] \right\rangle (\omega, \mathbf{p}) 
\overset{\mathcal{O}(\lambda^0)}{=} 2\pi\, \mathrm{sgn}(\omega)\, \delta\!\left(\omega^2 - (\omega^{\mathrm{rel}}_\mathbf{p})^2\right) \delta_{ab}.  
\label{eq:spectral_function_pert}
\end{equation}
From higher orders in the coupling, the spectral function would receive corrections leading to a mass-shift and non-zero width of the spectral function encoding ``off-shell'' contributions to processes. However, since they are of higher order in the perturbative power counting for the collision term, we do not consider them here. Since there are only elastic collisions contributing to this order, particle number is artificially conserved. Inelastic processes can also be taken into account by going to higher order in the coupling~\cite{Moore:2015adu}. Starting from a general out-of-equilibrium state such inelastic processes are relevant to describe the approach to thermal equilibrium at late times, since otherwise a thermal distribution with chemical potential for particle number would appear even in the absence of a conserved number. Essentially, neglecting inelastic processes limits the time until which the approximation can be applied \cite{Arrizabalaga:2005tf}. For the purposes of this section going beyond the given order is not necessary.\footnote{For instance, for a relativistic scalar field theory describing interacting pions in the context of heavy-ion collisions, conditions for a conserved particle number density and the time interval where this approximation can be trusted have been discussed in Refs.~\cite{Voskresensky:1994uz,Voskresensky:1996ur,Kolomeitsev:2017wjd}.} 

Similarly, for the non-relativistic dispersion $\omega_\mathbf{p} = |\mathbf{p}|^2/2m$ at lowest perturbative order, the collision term for the kinetic equation of the theory with action (\ref{eq:classicalaction_nr}) becomes~\cite{Nazarenko:2011} 
\begin{align}
&C\nrsup[f](t,\mathbf{p}) = \int_\mathbf{l,q,r} 2g^2\, \FFunc[f](t,\mathbf{p,l,q,r}) \nonumber\\
& \times (2\pi)^{4}\delta^{(3)}(\mathbf{p+l-q-r})\;\delta \left(\omega_\mathbf{p} + \omega_\mathbf{l} - \omega_\mathbf{q} - \omega_\mathbf{r} \right)\,.
\label{eq_collision_integral_nr_perp}
\end{align}
The quadratic dispersion relation at this order can also be viewed as arising from the low-momentum limit of the above relativistic collision integral (\ref{eq_collision_integral_rel_perp}) \cite{Semikoz:1995rd,Micha:2004bv}.
We note that taking into account subleading corrections for non-relativistic theories one generally has a Bogoliubov dispersion relation with a quadratic dispersion at higher and a linear dispersion at lower momenta if a Bose-Einstein condensate exists \cite{Nazarenko:2011}, which we do not consider here. 

The perturbative power counting for $\lambda \ll 1$ leading to (\ref{eq_collision_integral_rel_perp}), or (\ref{eq_collision_integral_nr_perp}) for $\zeta \ll 1$, with elastic $2 \leftrightarrow 2$ scatterings as in (\ref{eq_functional_F}) assumes that the relevant occupancies $f_\mathbf{p}$ for typical momenta are not too high. More precisely, only for $f_\mathbf{p} \ll 1/\lambda$ in the relativistic, or $f_\mathbf{p} \ll 1/\zeta$ in the non-relativistic case, the higher-order corrections are parametrically small. Before we discuss this issue in more detail below in section~\ref{sec_applicability}, we will introduce in the following an alternative kinetic description based on a large-$N$ expansion of the underlying quantum field theory.

\section{Large-$N$ kinetic theory}
\label{sec:largeN}

The standard Boltzmann equation described in the last section is based on a weak-coupling expansion, which restricts the range of validity of the kinetic theory to perturbative problems. However, for the $N$-component field theory an alternative kinetic description with an extended range of validity may be derived based on a non-perturbative expansion in $N$. For details about its derivation from the underlying quantum field theory we refer to Refs.~\cite{Orioli:2015dxa,Berges:2015kfa}. Here we give the relevant expressions that are used below to solve the large-$N$ kinetic equation. 

At large $N$ the classical action (\ref{eq:classicalaction}) scales proportional to $N$ employing $\varphi_a \varphi_a \sim N$. Genuine quantum corrections due to scatterings appear at subleading orders in a large-$N$ expansion, i.e., they are down by factors of $1/N$ compared to classical contributions~\cite{Aarts:2002dj,Scheppach:2009wu}. 
In particular, the spectral function at leading order (LO) in a large-$N$ expansion reads
\begin{equation}
\left\langle \left[ \varphi_a, \varphi_b \right] \right\rangle (\omega, \mathbf{p}) \overset{{\rm LO}\, {\rm large}\,N}{=} 2\pi\, \mathrm{sgn}(\omega)\, \delta\!\left(\omega^2 - (\omega^{\mathrm{rel}}_\mathbf{p})^2\right) \delta_{ab} ,
\label{eq:spectral_function_NLO}
\end{equation}
where, in contrast to the lowest-order perturbative Eq.~\eqref{eq:reldispersion}, the dispersion for the relativistic theory now contains an effective mass term $M^2$:
\begin{equation}
\omega\relsup_\mathbf{p} = \sqrt{M^2 + \mathbf{p}^2}  \, .
\end{equation}
At LO the effective mass term is given by the gap equation~\cite{Aarts:2002dj,Orioli:2015dxa}
\begin{align}
 M^2 = m^2 + \frac{\lambda}{6}\int_{\mathbf{p}} \frac{f(t_0,\mathbf{p})}{\omega^{\mathrm{rel}}_\mathbf{p}} 
 \label{eq_eff_mass_def}
\end{align}
evaluated at some given time $t_0$. The mass term is constant at lowest order in the gradient expansion underlying kinetic descriptions~\cite{Berges:2005md}.

We will describe in the following that at NLO in the large-$N$ expansion, there is a well-defined effective kinetic description in terms of scatterings between quasiparticles. Similar to the previous section~\ref{sec:pertkin}, we start by considering the relativistic theory and extend the discussion to the non-relativistic case in the end. 

%%
%% different channels of the resummed vertex
%%%%%%%%%%%%%%%%%%%%%%%%%%%%%%%%%%%%%%%%%%%%%%%%%%%%%%%%%%%%%%%%%%%%%%%%%%%%%%%%%%%%%%%%%%%%%%%%%%%%
\begin{figure}[t]
	\centering
	\vspace*{-6ex}
	\includegraphics[scale=0.32]{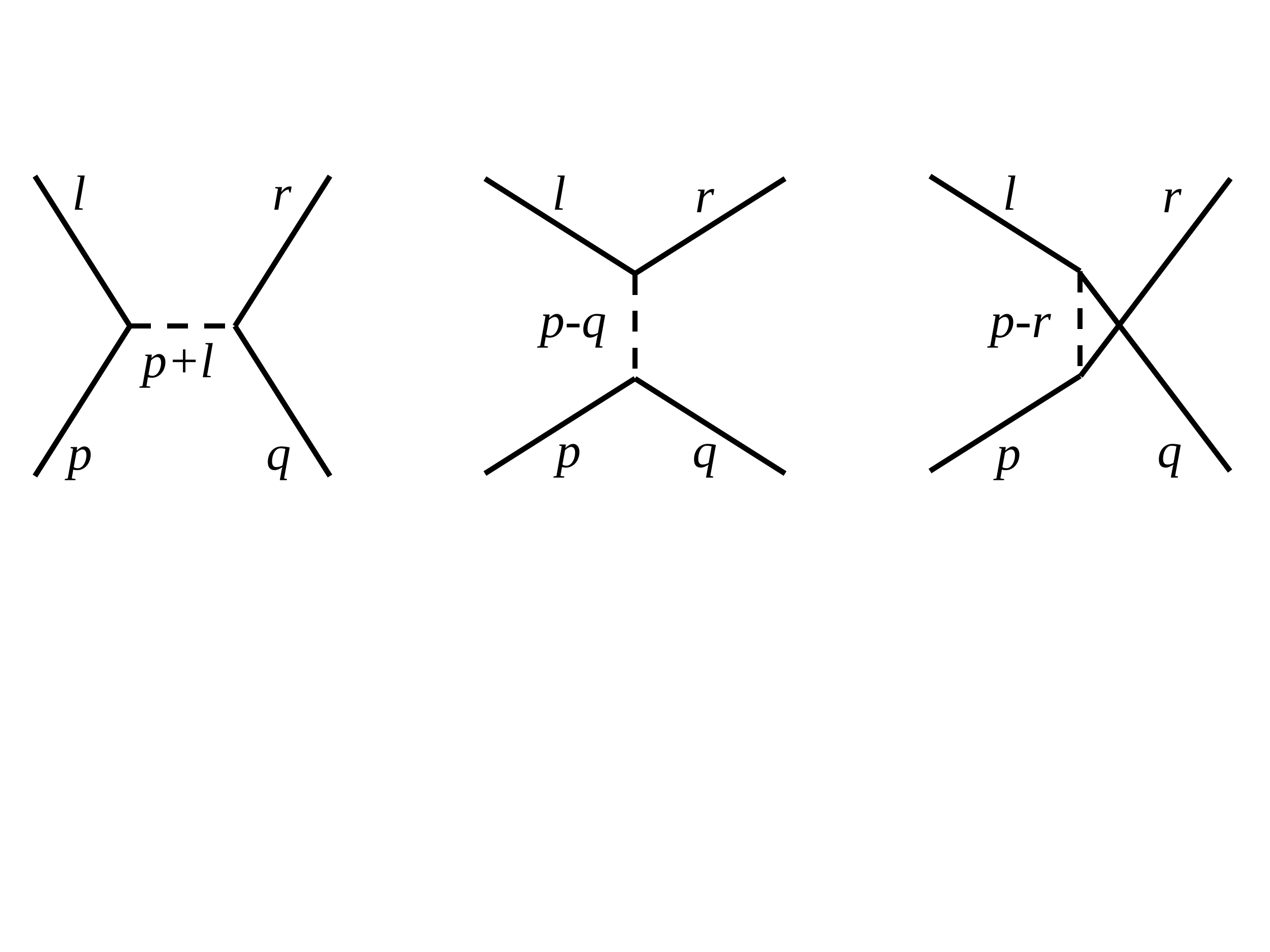}
	\vspace*{-20ex}
	\caption{Scattering processes at next-to-leading order in the large-$N$ expansion, which are mediated by an effective interaction. While the solid lines represent particles with given four-momenta, the dashed line represents the function (\ref{eq:auxiliaryprop}) in Fourier space describing $s$-, $t$- and $u$-channel exchange.}
	\label{fig:stu}
\end{figure}
%%%%%%%%%%%%%%%%%%%%%%%%%%%%%%%%%%%%%%%%%%%%%%%%%%%%%%%%%%%%%%%%%%%%%%%%%%%%%%%%%%%%%%%%%%%%%%%%%%%%
%%
%%

In order to discuss subleading corrections in the $1/N$ expansion, it is convenient to employ the auxiliary field formulation of the same model~\cite{Aarts:2002dj}. For this purpose, we rewrite the original action (\ref{eq:classicalaction}) by introducing an auxiliary field $\chi(x)$ as
\begin{equation}
S[\varphi,\chi] = - \! \int_{t,\mathbf{x}} \! \left[ \frac{1}{2} \varphi_a \! \left( \Box + m^2 \right) \! \varphi_a - \frac{3N}{2\lambda} \chi^2 + \frac{1}{2} \chi \varphi_a \varphi_a\right] \! .
\label{eq:classicalaction_aux}
\end{equation}
Integrating out $\chi$ in the defining functional integral yields the original action, and from the Heisenberg equations of motion one sees that the auxiliary field represents the composite operator
\begin{equation}
\chi(x) = \frac{\lambda}{6N}\varphi_a(x) \varphi_a(x) \, .
\label{eq:chi}
\end{equation}
While the auxiliary field is not a dynamical degree of freedom, it can be used to conveniently express scattering corrections in terms of the expectation value
\begin{equation}
D(x-y) \equiv \langle \chi(x) \chi(y) \rangle - \langle \chi(x) \rangle \langle \chi(y) \rangle \, .
\label{eq:auxiliaryprop}
\end{equation}
Since $\chi$ represents a two-point function according to (\ref{eq:chi}), the function $D(x,y)$ encodes a four-point function or vertex. Specifically, at NLO in the $1/N$ expansion, scatterings are mediated by (\ref{eq:auxiliaryprop})~\cite{Aarts:2002dj}. This is indicated in Fig.~\ref{fig:stu}, where dashed lines represent the two-point function (\ref{eq:auxiliaryprop}) in Fourier space. The modified vertex at NLO is shown for scatterings in the $s$-, $t$, and $u$-channel, respectively. 

Accordingly, the effective kinetic equation at NLO is given by the same kinetic equation (\ref{eq_effective_Boltzmann_equation}), however, with the different collision term~\cite{Orioli:2015dxa}
\begin{align}
&C\relsup_\mathrm{NLO}[f](t,\mathbf{p}) = \int_\mathbf{l,q,r}\!\!\!\! \frac{\lambda^2_{\rm eff}(t,{\mathbf p},{\mathbf l},{\mathbf q},{\mathbf r})}{6 N} \, \FFunc[f](t,\mathbf{p,l,q,r}) \nonumber\\
& \times (2\pi)^4\delta^{(3)}(\mathbf{p+l-q-r})\frac{\delta(\omega\relsup_\mathbf{p} + \omega\relsup_\mathbf{l} - \omega\relsup_\mathbf{q} - \omega\relsup_\mathbf{r} )}{2 \omega\relsup_\mathbf{p}\,2 \omega\relsup_\mathbf{l}\,2 \omega\relsup_\mathbf{q}\,2 \omega\relsup_\mathbf{r}}.
\label{eq_collision_integral_rel_perp_NLO}
\end{align}
In the derivation of the collision integral, the LO expression of the spectral function (\ref{eq:spectral_function_NLO}) is used since its subleading corrections in $1/N$ would result in subleading corrections of the collision integral, which are part of the large-$N$ kinetic theory at next-to-next-to-leading order (NNLO), and are thus omitted. 
The time- and momentum-dependent effective coupling function
\begin{eqnarray}
\lambda^2_{\rm eff}(t,{\mathbf p},{\mathbf l},{\mathbf q},{\mathbf r}) &\equiv& \frac{\lambda^2}{3}
\left[ \frac{1}{| 1 + \Pi_{R}^{\mathrm{rel}} (t,\omega\relsup_{\mathbf p} + \omega\relsup_{\mathbf l},{\mathbf p}+{\mathbf l})|^2} \right. \nonumber\\
&+& \frac{1}{| 1 + \Pi_{R}^{\mathrm{rel}} (t,\omega\relsup_{\mathbf p} - \omega\relsup_{\mathbf q},{\mathbf p}-{\mathbf q})|^2} 
\nonumber\\
&+&  \!\left. \frac{1}{| 1 + \Pi_{R}^{\mathrm{rel}} (t,\omega\relsup_{\mathbf p} - \omega\relsup_{\mathbf r},{\mathbf p}-{\mathbf r})|^2} \right] \quad
\label{eq:leffrel}
\end{eqnarray}
incorporates the vertex corrections for the different scattering channels according to Fig.~\ref{fig:stu}. The appearance of the renormalized ``one-loop'' retarded self-energy
\begin{eqnarray}
&& \Pi_R^{\mathrm{rel}}(t,\omega,\mathbf p) \, = \, \frac{\lambda}{12} \int_\mathbf{q} \, \frac{f(t,\mathbf p-\mathbf q)}{\omega\relsup_{\mathbf q}\, \omega\relsup_{\mathbf p-\mathbf q}} \nonumber\\
&& \quad \times \bigg[ \frac{1}{\omega\relsup_{\mathbf q}+\omega\relsup_{\mathbf p-\mathbf q}-\omega-i\epsilon} + \frac{1}{\omega\relsup_{\mathbf q}-\omega\relsup_{\mathbf p-\mathbf q}-\omega-i\epsilon}  \nonumber\\
&& \quad  + \;\;\frac{1}{\omega\relsup_{\mathbf q}-\omega\relsup_{\mathbf p-\mathbf q}+\omega+i\epsilon} + \frac{1}{\omega\relsup_{\mathbf q}+\omega\relsup_{\mathbf p-\mathbf q}+\omega+i\epsilon} \bigg]   \qquad
\label{eq:pir_onshell_rel}
\end{eqnarray}
in the denominator of Eq.~(\ref{eq:leffrel}) is the result of a geometric series summation of an infinite number of scattering processes at NLO in the large-$N$ expansion~\cite{Berges:2001fi}. We emphasize that $\Pi_R^{\mathrm{rel}}$, and thus also $\lambda^2_{\rm eff}$, is time-dependent since it depends on the evolving distribution function.

From (\ref{eq:leffrel}) one observes that for $|\Pi_{R}^{\mathrm{rel}}| \ll 1$, which is the case for weak enough coupling ($\lambda \ll 1$) and not too large typical occupancies ($f \ll 1/\lambda$), the vertex corrections encoded in the momentum-dependent effective coupling become irrelevant such that $\lambda^2_{\rm eff} \simeq \lambda^2$. In this case the collision term (\ref{eq_collision_integral_rel_perp_NLO}) essentially describes standard perturbative $2 \leftrightarrow 2$ scatterings, however, at large $N$.\footnote{\label{foot:prefactor_rel}The prefactor $(N+2)/(6N^2)$ in Eq.~(\ref{eq_collision_integral_rel_perp}) becomes $1/(6N)$ at NLO in the large-$N$ expansion.} In contrast, for high characteristic occupancies with $f \sim 1/\lambda$ the collision term (\ref{eq_collision_integral_rel_perp_NLO}) can be strongly modified if $\Pi_{R}^{\mathrm{rel}}$ starts to become of order one. We will discuss the corresponding behavior of the effective coupling in more detail in the following sections, for which we will introduce the non-relativistic effective kinetic equation relevant at low momenta below. 

Similar to the lowest-order perturbative kinetic equation of Sec.~\ref{sec:pertkin}, the large-$N$ kinetic theory at NLO only involves elastic scattering processes. The lack of inelastic processes implies conservation of the particle number density $n = \int_\mathbf{p} f(t,\mathbf{p}) = \mathrm{const}$, which follows from the kinetic equation \eqref{eq_effective_Boltzmann_equation} and $\int_\mathbf{p} C\relsup_\mathrm{NLO}[f](t,\mathbf{p}) = 0$. Taking into account inelastic processes is possible by going beyond NLO, which is however beyond the scope of the present study that aims to provide a kinetic description of number conserving dynamics near nonthermal fixed points~\cite{Orioli:2015dxa,Moore:2015adu,Berges:2017ldx}. Moreover, here the condensate formation time diverges with volume $\sim V^{1/\alpha}$ with positive scaling exponent $\alpha$ as shown in lattice simulations \cite{Orioli:2015dxa}. Hence, no emergence of a condensate is expected within a finite time for the infinite volume considered, which is consistent with the self-similar evolution that we will observe in numerical calculations of the large-$N$ kinetic theory in Sec.~\ref{sec_solution}. Therefore, we will consider the collision integral \eqref{eq_collision_integral_rel_perp_NLO} without including a condensate in the following. Further discussions on the distinction between the perturbative and non-perturbative regimes can be found in \cite{Chantesana:2018qsb}. 

To simplify the following discussion and to make the connection to ultracold atoms, we will restrict ourselves to momenta below the (effective) mass scale $|\mathbf{p}| \ll M$ and describe the dynamics in terms of a non-relativistic quantum field theory. 
%Because of particle number conservation, the effective mass $M$ is approximately constant when the integral in Eq.~\eqref{eq_eff_mass_def} is dominated by low momenta $|\mathbf{p}| \ll M$. This is consistent with the non-relativistic limit of the large-$N$ kinetic theory and we will consider $M$ as a constant parameter in the following. 
Note that also relativistic theories with $m = 0$ but $M > 0$ may be described by the non-relativistic limit for small momenta. To be consistent with the non-relativistic theory defined by \eqref{eq:classicalaction_nr}, we will use the symbol $m$ for the mass.

Following along the lines of section \ref{sec:pertkin}, we consider first the case $N=2$ to illustrate the effective kinetic equation for a non-relativistic complex scalar field, i.e., with two real field components. The case of general $N$ then proceeds accordingly~\cite{Gasenzer:2005ze}. 

For the quadratic dispersion relation one obtains~\cite{Orioli:2015dxa}
\begin{align}
&C\nrsup_\mathrm{NLO}[f](t,\mathbf{p})  \nonumber\\
& = \int_\mathbf{l,q,r} g_{\mathrm{eff}}^2[f](t,\omega_\mathbf{p}-\omega_\mathbf{q},\mathbf{p-q})\, \FFunc[f](t,\mathbf{p,l,q,r}) \nonumber\\
& \times (2\pi)^{4}\delta^{(3)}(\mathbf{p+l-q-r})\;\delta \left(\omega_\mathbf{p} + \omega_\mathbf{l} - \omega_\mathbf{q} - \omega_\mathbf{r} \right)
\label{eq_collision_integral_raw}
.
\end{align}
The effective coupling in the collision integral reads
\begin{align}
g_{\mathrm{eff}}^2[f](t,\omega,\mathbf{\pMq}) = \frac{g^2}{\left| 1 + \PiR\left( t, \omega, \mathbf{\pMq} \right) \right|^2}\,,
\label{eq_geff}
\end{align}
with the ``one-loop'' retarded self-energy
\begin{align}
&\Pi_R\left( t, \omega, \mathbf{\pMq} \right) = \lim_{\epsilon \to 0^+} g \int_\mathbf{k} \, f(t, \mathbf{\pMq-k}) \nonumber\\
&\times \left[ \frac{1}{\omega_\mathbf{k} - \omega_\mathbf{\pMq-k} - \omega - i\epsilon} + \frac{1}{\omega_\mathbf{k} - \omega_\mathbf{\pMq-k} + \omega + i\epsilon} \right],
\label{eq_1loop_ret_self_energy_raw}
\end{align}
and the momentum difference $\mathbf{\pMq} = \mathbf{p-q}$. As for the relativistic theory, the collision integral (\ref{eq_collision_integral_raw}) reduces to its perturbative expression\footnote{The factor of 2 difference between the large-$N$ NLO expression (\ref{eq_collision_integral_raw}) and the perturbative collision integral in (\ref{eq_collision_integral_nr_perp}) is due to an omitted term that is of order NNLO. Note that a similar modification is found for the relativistic theory, as commented in footnote \ref{foot:prefactor_rel}.} for small $|\PiR| \ll 1$.

For later use, it is helpful to further evaluate the expressions for $C\nrsup_\mathrm{NLO}$ and $\Pi_R$. Using magnitudes of momenta $p = |\mathbf{p}|$, and similar for $q$, $k$ and $\pMq$, we find for an isotropic system (details are given in App.~\ref{sec_preparation}) 
\begin{align}
\label{eq_collision_integral_massaged}
&C[f](t,p)
=
\frac{m}{32 \pi^3 p} \int_{0}^{\infty} \mathrm{d} q \, q
\int_{\left|p-q\right|}^{p+q} \text{d} \pMq
\nonumber\\
&\times g_\mathrm{eff}^2[f]( t,\omega_p - \omega_q, \pMq )
\nonumber\\
&\times \int_{\mathrm{max}\left(\pMq, \frac{\left|p^2-q^2\right|}{\pMq}\right)}^{\infty} \mathrm{d} u 
\left( u - \frac{\left(p^2 - q^2 \right)^2}{u^3} \right)
\nonumber \\
&\times \FFunc[f]\left( t, p, \frac{1}{2}\left( u - \frac{p^2 - q^2}{u} \right), q, \frac{1}{2}\left( u + \frac{p^2 - q^2}{u} \right) \right)
\end{align}
and 
\begin{align}
&\Pi_R(t,\omega,\pMq)
= \frac{mg}{(2\pi)^2 \pMq} \Bigg\{ \int_{0}^{\infty} \mathrm d k \, k \, f(t,k) \nonumber\\
&\times \log\left| \frac{ \left(k + \frac{\pMq}{2}\right)^2 - \frac{m^2 \omega^2}{\pMq^2} }{ \left(k - \frac{\pMq}{2}\right)^2 - \frac{m^2 \omega^2}{\pMq^2} } \right| + i \pi \int_{\frac{|\pMq^2 - 2m\omega|}{2\pMq}}^{\frac{|\pMq^2 + 2m\omega|}{2\pMq}} \mathrm d k \, k \, f(t,k) \Bigg\},
\label{eq_1loop_ret_self_energy_massaged}
\end{align}
where we have dropped the labels of $C\nrsup_\mathrm{NLO}$ to shorten the notation. Accordingly, the effective kinetic equation (\ref{eq_effective_Boltzmann_equation}) depends on time $t$ and the magnitude of the momentum $p$.

\section{Behavior of the large-$N$ resummed effective vertex}
\label{sec_scaling_effective_coupling}

To discuss the extended range of validity of large-$N$ kinetic theory, we first consider the effective coupling $g_\mathrm{eff}^2$ appearing in the collision integral (\ref{eq_collision_integral_massaged}), which is a function of the difference in energies of the in- and outgoing particles $\omega = \omega_p - \omega_q = (p^2 - q^2)/2m$ and of the magnitude of the momentum change $\pMq = |\mathbf{p-q}|$ in a scattering event. It is beneficial to consider limiting cases of its arguments to analyze its behavior. We distinguish three typical collision scenarios. In the first case, the momentum of a particle within a collision is strongly decelerated so that $p \gg q$ and thus $2m\omega \approx p^2 \approx \pMq^2$. Similarly, $q \gg p$ leads to the same effective coupling because $g_\mathrm{eff}^2[f](t,\omega,\pMq)$ is symmetric in $\omega$. To ease the discussion we will therefore use $\omega \geq 0$ in the following. In the second case, the magnitude of the momentum of the particle undergoing the collision stays at the same order $p \sim q$ while it changes its direction. The nearly collinear regime is discussed separately and constitutes the third scenario. 

The effective coupling in~\eqref{eq_geff} can be calculated from the retarded self-energy $\PiR(t,\omega, \pMq)$ given in~(\ref{eq_1loop_ret_self_energy_massaged}). For the distribution function $f(t,k)$ that enters the integrals in $\PiR$, we assume that for $l = 0,1,2$ integrals of the form
\begin{align}
	\int dk\, k^{l} f(t,k) \sim K^{l+1} f(t,K)
	\label{eq_parametric_ints}
\end{align}
are dominated at the possibly time-dependent momentum scale $K$, if it lies within the integration limits. This scale $K$ can be defined as the momentum where $k^2\, f(t,k)$ is maximal
\begin{align}
 \label{eq_def_scale_K}
 \left.k^2\,f(t,k)\right|_{k=K} = \mathrm{max} \left( k^2\,f(t,k) \right),
 \end{align}
such that it provides the dominant contributions to the particle number density
\begin{align}
 \label{eq_particle_number}
 n \propto \int \frac{\mathrm{d}^3 k}{(2 \pi)^3}\, f(t,k) \sim K^3\, f(t,K)\,.
\end{align}
For the integrals in (\ref{eq_parametric_ints}) to converge between the maximal limits of $0$ and $\infty$, the distribution function $f(t,k)$ should fall off faster than $k^{-3}$ at large momenta $k \ggSim K$ and decrease slower than $k^{-1}$ at low momenta $k \llSim K$, or not decrease there at all.

\subsection{Dispersive regime, $\pMq^2 \approx 2m\omega$}
\label{sec_disp_regime}

We start with the regime $p \gg q$. Then one has $\pMq \approx p$ and $\omega \approx \omega_p = p^2/2m \approx \pMq^2/2m$. The second relation states that the energy difference of in- and outgoing momenta $\omega$ follows the non-relativistic dispersion relation with momentum (difference) $\pMq$, and we refer to this regime as dispersive. The ``one-loop'' retarded self-energy~\eqref{eq_1loop_ret_self_energy_massaged} in this limit reads
\begin{align}
	\PiR\left(t,\frac{\pMq^2}{2m},\pMq\right)
	=\; &\frac{mg}{(2\pi)^2\, \pMq} \Bigg\{ \int_{0}^{\infty} \mathrm d k \, k \, f(t,k) \log\left| \frac{k + \pMq}{k - \pMq} \right| \nonumber\\
	& \qquad +i \pi \int_{0}^{\pMq} \mathrm d k \, k \, f(t,k) \Bigg\}.
\end{align}
Its real part involves an integration over all momenta and we can use \eqref{eq_parametric_ints} in the limiting cases of $\pMq \ggSim K$ and $\pMq \llSim K$. In the first case, the integrand is dominated at low momenta $k$ and we can approximate $\log(k + \pMq) - \log|k - \pMq| \approx 2k/\pMq + \mathcal{O}((k/\pMq)^3)$. Similarly, the second case leads to $\log(k + \pMq) - \log|k - \pMq| \approx 2\pMq/k + \mathcal{O}((\pMq/k)^3)$. Hence, the limiting expressions are 
\begin{align}
	\label{eq_pGGq_RePi_pGGK}
	\textrm{Re}\, \PiR\left(t,\frac{\pMq^2}{2m},\pMq\right)\; &\overset{\pMq \, \ggSim \, K}{\sim} m\, g\, K\, f(t,K)\; \frac{K^2}{\pMq^2}  \\
	\textrm{Re}\, \PiR\left(t,\frac{\pMq^2}{2m},\pMq\right)\; &\overset{\pMq \, \llSim \, K}{\sim} m\, g\, K\, f(t,K)\,.
	\label{eq_pGGq_RePi_pLLK}
\end{align}
For the imaginary part, large and small ingoing momenta $\pMq \gtrsim K$ and $\pMq \lesssim K$ lead to the expressions
\begin{align}
	\label{eq_pGGq_ImPi_pGGK}
	\textrm{Im}\, \PiR\left(t,\frac{\pMq^2}{2m},\pMq\right)\; &\overset{\pMq \, \gtrsim \, K}{\sim} m\, g\, K\, f(t,K)\; \frac{K}{\pMq}  \\
	\textrm{Im}\, \PiR\left(t,\frac{\pMq^2}{2m},\pMq\right)\; &\overset{\pMq \, \lesssim \, K}{\sim} m\, g\, \pMq\, f(t,\pMq)\,.
	\label{eq_pGGq_ImPi_pLLK}
\end{align}
In (\ref{eq_pGGq_ImPi_pLLK}) we used that $k f(t,k)$ should be a growing function at low momenta to be consistent with (\ref{eq_parametric_ints}).

To get the corresponding limiting expressions for the effective coupling, we first assume that for typical soft momenta $K$ the occupation number $f(t,K)$ is sufficiently large such that for the considered momenta $\pMq$, the $1$ in the denominator of $g_{\mathrm{eff}}^2$ in (\ref{eq_geff}) can be neglected, and the effective coupling reads $g_{\mathrm{eff}}^2 \approx g^2\left( (\textrm{Re}\, \PiR)^2 + (\textrm{Im}\, \PiR)^{2} \right)^{-1}$. Since $\pMq\, f(t,\pMq)$ is limited by $K\, f(t,K)$, the real part (\ref{eq_pGGq_RePi_pLLK}) dominates at low momenta $\pMq \, \llSim \, K$. On the other hand, the imaginary part (\ref{eq_pGGq_ImPi_pGGK}) decreases slower than the real part at high momenta $\pMq \, \ggSim \, K$, and is thus larger. With this, the effective coupling is parametrically
\begin{align}
	\label{eq_pGGq_geff_pGGK}
	g_{\mathrm{eff}}^2[f]\left(t,\frac{\pMq^2}{2m},\pMq\right)\; &\overset{\pMq \, \ggSim \, K}{\sim} \frac{1}{(m\, K\, f(t,K))^2} \; \frac{\pMq^2}{K^2}  \\
	g_{\mathrm{eff}}^2[f]\left(t,\frac{\pMq^2}{2m},\pMq\right)\; &\overset{\pMq \, \llSim \, K}{\sim} \frac{1}{(m\, K\, f(t,K))^2}\,.
	\label{eq_pGGq_geff_pLLK}
\end{align}
Accordingly, it is constant below $K$ and follows the power law $\pMq^2$ beyond $K$. We note that at even larger momenta, it becomes constant $\simeq g^2$ when the $+1$ in the denominator of its definition becomes important.

As noted above, the regime $q \gg p$ leads to the same expressions (\ref{eq_pGGq_geff_pGGK}), (\ref{eq_pGGq_geff_pLLK}), with $\pMq \approx q$. The frequency argument gets a minus sign $-q^2/2m$, which does not change the values of $g^2_\mathrm{eff}$ because of its symmetry.

\subsection{Momentum-dominated regime, $\pMq^2 \ggSim 2m\omega$}
\label{sec_momentum_dominated}

In the momentum-dominated regime, we consider the situation when in- and outgoing momenta are of the same order $p \sim q$. For further simplifications, we also assume that the frequency difference is small $\omega = \omega_p - \omega_q \llSim \pMq^2/2m$. This occurs for most of the scattering angles $\cos \theta_{\mathbf{pq}} = \mathbf{pq}/pq$, since this assumption is equivalent to the condition $\cos \theta_{\mathbf{pq}} \llSim \min(q/p,p/q)$. The remaining case of nearly collinear collisions $\cos \theta_{\mathbf{pq}} \sim 1$ will be discussed in Sec.~\ref{sec_collinear}.

In the considered regime, the imaginary and real parts of the ``one-loop'' retarded self-energy (\ref{eq_1loop_ret_self_energy_massaged}) become
\begin{align}
	\textrm{Re}\, \PiR\left(t,\omega,\pMq\right) \,&\approx \frac{mg}{(2\pi)^2 \pMq} \int_{0}^{\infty} \mathrm d k \, k \, f(t,k) \, 2\log\left| \frac{2k + \pMq}{2k - \pMq} \right| \\
	\textrm{Im}\, \PiR\left(t,\omega,\pMq\right) \,&\approx \frac{mg}{4\pi} \;\frac{\pMq}{2}\, f\left(t,\frac{\pMq}{2} \right)\; \frac{2m\,\omega}{\pMq^2}\,.
\end{align}
Expanding the logarithm of the real part and approximating the integral as in (\ref{eq_parametric_ints}), one arrives at similar expressions as in the dispersive case
\begin{align}
	\label{eq_pSIMq_RePi_pGGK}
	\textrm{Re}\, \PiR\left(t,\omega,\pMq \right)\; &\overset{\pMq \, \ggSim \, 2K}{\sim} m\, g\, K\, f(t,K)\; \frac{(2K)^2}{\pMq^2}  \\
	\textrm{Re}\, \PiR\left(t,\omega,\pMq \right)\; &\overset{\pMq \, \llSim \, 2K}{\sim} m\, g\, K\, f(t,K)\,.
	\label{eq_pSIMq_RePi_pLLK}
\end{align}
A closer look on the logarithm of the original expression in (\ref{eq_1loop_ret_self_energy_massaged}) reveals that if $2 m \omega \llSim 2K \pMq$ is satisfied, the estimate for lower momenta (\ref{eq_pSIMq_RePi_pLLK}) is even valid in the collinear regime where $2 m \omega$ exceeds $\pMq^2$. Hence, the full range of validity for (\ref{eq_pSIMq_RePi_pLLK}) is $2K \ggSim \pMq \ggSim 2m\omega / 2K$, which may only hold if $(2K)^2 \ggSim 2m\omega$. Otherwise, for $(2K)^2 \llSim 2m\omega$, no region with the value~(\ref{eq_pSIMq_RePi_pLLK}) exists. 

To compute the effective coupling, we again assume large occupation numbers and thus neglect the $1$ in the denominator in (\ref{eq_geff}), which yields $g_{\mathrm{eff}}^2 \approx g^2\left( (\textrm{Re}\, \PiR)^2 + (\textrm{Im}\, \PiR)^{2} \right)^{-1}$. For both small and large momenta $\pMq$, the real part is larger than the imaginary part $\textrm{Re}\, \PiR \ggSim \textrm{Im}\, \PiR$. This follows from $2m\omega \llSim \pMq^2$ and, for small momenta $\pMq / 2 \llSim K$, from $(\pMq / 2)\, f(t,\pMq / 2) \llSim K\, f(t,K)$ while for larger momenta $\pMq / 2 \ggSim K$ it results from $(\pMq/2)^3\, f(t,\pMq/2) \llSim K^3\, f(t,K)$, that are both requirements for $f(t,k)$ and were formulated below Eq.~(\ref{eq_parametric_ints}). Hence, the effective coupling parametrically follows
\begin{align}
	\label{eq_pSimq_geff_pGGK}
	g_{\mathrm{eff}}^2[f]\left(t, \omega, \pMq \right)\; &\overset{\pMq \, \ggSim \, 2K}{\sim} \frac{1}{(m\, K\, f(t,K))^2} \; \frac{\pMq^4}{(2K)^4}  \\
	g_{\mathrm{eff}}^2[f]\left(t, \omega, \pMq \right)\; &\overset{\pMq \, \llSim \, 2K}{\sim} \frac{1}{(m\, K\, f(t,K))^2}\,.
	\label{eq_pSimq_geff_pLLK}
\end{align}
The main difference from the dispersive regime is the steep power law $\pMq^4$ at large momenta. Interestingly, the transition between small and large momentum expressions proceeds at the slightly larger scale $2K$.

\subsection{Collinear regime, $\pMq^2 \llSim 2m\omega$}
\label{sec_collinear}

The remaining case is when in- and outgoing momenta are nearly collinear $\cos \theta_{\mathbf{pq}} \sim 1$, i.e. the case where $\pMq^2 \llSim 2m\omega$. The logarithm appearing in $\textrm{Re}\, \PiR$ in (\ref{eq_1loop_ret_self_energy_massaged}) can be written as
\begin{align}
	\log\left| 1 - \frac{\pMq^2 (2k + \pMq)^2}{(2 m \omega)^2} \right| \, &- \log \left| 1 - \frac{\pMq^2 (2k - \pMq)^2}{(2 m \omega)^2} \right| \nonumber \\
	&\approx -4 \,\frac{\pMq^4}{(2 m \omega)^2}\, \frac{2k}{\pMq}\,,
\end{align}
where we have expanded it in the second line. Assuming that the integral is dominated at momenta $k \sim K$ as in the cases above, this expansion is justified for large momenta $\pMq \, \ggSim \, 2K$ while the condition $2 m \omega \ggSim 2K \pMq$ is additionally required at low momenta $\pMq \, \llSim \, 2K$. 

With this, we can readily estimate the real and imaginary parts of $\PiR$ as
\begin{align}
	\label{eq_pSIMq_llW_RePi}
	\textrm{Re}\, \PiR\left(t,\omega,\pMq\right) \,&\sim - m\, g\, K\, f(t,K)\; \frac{(2K)^2}{(\Pinv)^2}  \\
	\textrm{Im}\, \PiR\left(t,\omega,\pMq\right) \,&\approx \frac{mg}{4\pi} \;\frac{\Pinv}{2}\, f\left(t,\frac{\Pinv}{2} \right)\,,
	\label{eq_pSIMq_llW_ImPi}
\end{align}
where we have introduced the inverse momentum $\Pinv = 2 m \omega / \pMq$. Recall that for the real part, the remaining situation of $2 m \omega \llSim 2K \pMq$ for low momenta $\pMq \, \llSim \, 2K$ in the collinear regime has been discussed in Sec.~\ref{sec_momentum_dominated}, where it led to the expression (\ref{eq_pSIMq_RePi_pLLK}). 

From this, we can compute the effective coupling in the collinear regime. Interestingly, the real part is negative and $1 + \textrm{Re}\, \PiR$ in the denominator of the effective coupling (\ref{eq_geff}) may become zero, leading to a resonant increase of the coupling. Otherwise, if $f(t,K)$ is sufficiently large, the $1$ in the denominator of (\ref{eq_geff}) can again be neglected and the real and imaginary parts of $\PiR$ can be compared in order to estimate $g_{\mathrm{eff}}^2 \approx g^2\left( (\textrm{Re}\, \PiR)^2 + (\textrm{Im}\, \PiR)^{2} \right)^{-1}$. At first, we consider $\Pinv \ggSim 2K$. This condition translates to $2 m \omega \ggSim 2K \pMq$ and corresponds to the real part as given by (\ref{eq_pSIMq_llW_RePi}). Using $(\Pinv)^3\, f(t,\Pinv) \llSim K^3\, f(t,K)$ for large momenta $\Pinv$, one finds $\textrm{Im}\, \PiR \llSim |\textrm{Re}\, \PiR|$. Similarly, the case $\Pinv \llSim 2K$ translates to $2 m \omega \llSim 2K \pMq$ and the real part is then given by (\ref{eq_pSIMq_RePi_pLLK}). With $\Pinv\, f(t,\Pinv) \llSim K\, f(t,K)$ for low momenta, one again finds $\textrm{Im}\, \PiR \llSim |\textrm{Re}\, \PiR|$. Thus, as in the momentum-dominated regime, the real part dominates the effective coupling. At low momenta $\pMq \, \llSim \, 2K$ it leads to (\ref{eq_pSimq_geff_pLLK}) for $2 m \omega \llSim 2K \pMq$ while in all other cases in the collinear regime, one has
\begin{align}
	g_{\mathrm{eff}}^2[f]\left(t, \omega, \pMq \right)\; &\sim \frac{1}{(m\, K\, f(t,K))^2} \; \frac{(2m\omega)^4}{(2K \pMq)^4}\,.
	\label{eq_pSimq_llW_geff}
\end{align}
Hence, different from the momentum-dominated regime, the effective coupling decreases here as $\pMq^{-4}$.

%%
%% effective coupling for omega = p^2 / 2m and = 0
%%%%%%%%%%%%%%%%%%%%%%%%%%%%%%%%%%%%%%%%%%%%%%%%%%%%%%%%%%%%%%%%%%%%%%%%%%%%%%%%%%%%%%%%%%%%%%%%%%%%
\begin{figure}[t]
	\centering
	\includegraphics[scale=0.23]{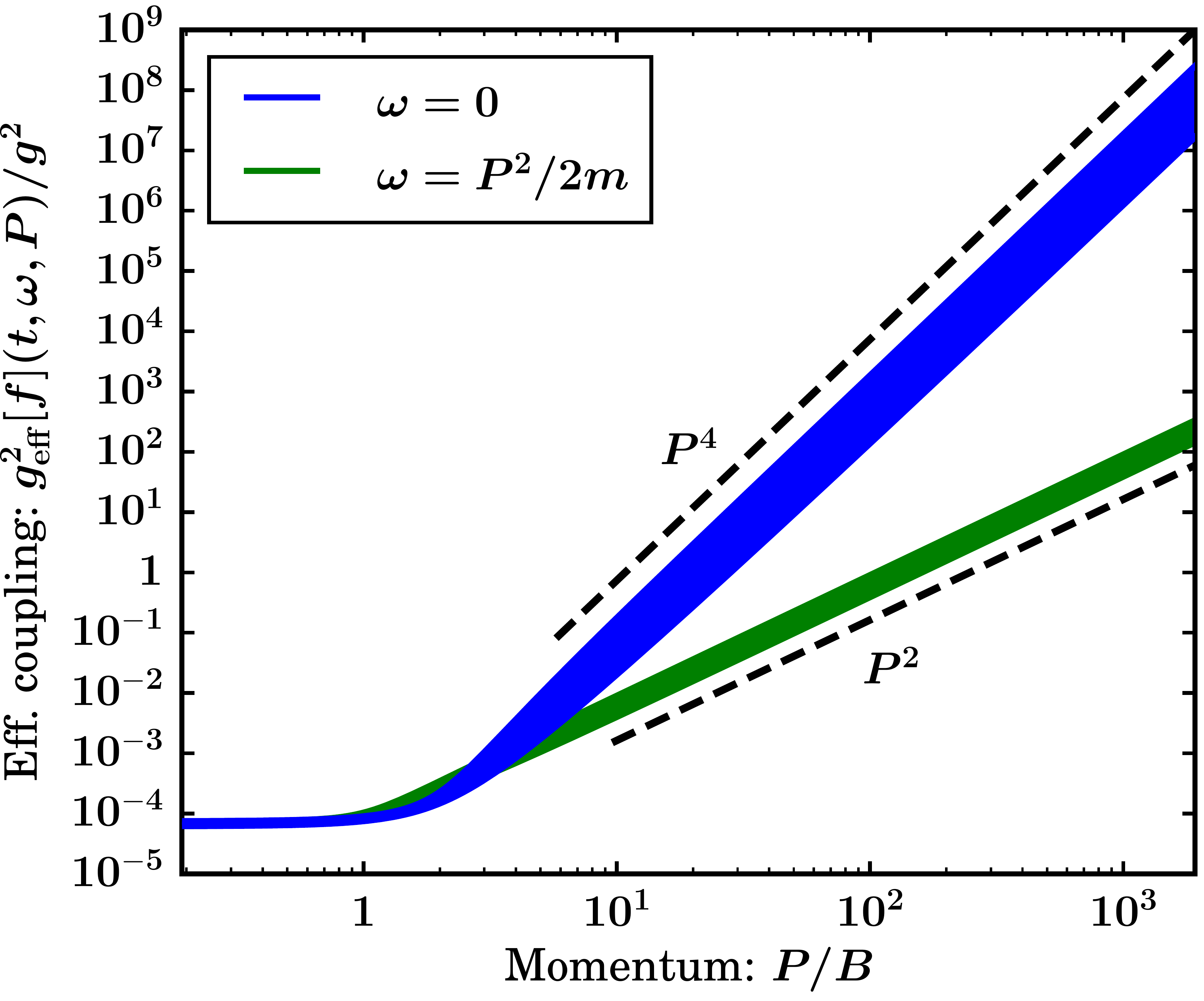}
	\caption{Effective coupling $g^2_\mathrm{eff}[f](t,\omega,\pMq)$ without $1$ in its denominator as a function of momentum for the momentum-dominated regime (blue line, $\omega = 0$) and for the dispersive regime (green line, $\omega = \pMq^2 / 2m$). The distribution function is chosen as in Eq.~\eqref{eq_fit_distribution_lattice} with parameters $A=84100 / (4m^2)$ and $B=0.0521 \times (2m)$. These values are taken from Ref.~\cite{Orioli:2015dxa}, where this form was fitted to the distribution function in a classical lattice simulation at time $t=300/(2m)$. Keeping $\kappa_< = 0$ fixed, we show bands for $3.5 \leq \kappa_> \leq 6$.}
	\label{fig_effective_coupling_omega_pp2m_and_0}
\end{figure}
%%%%%%%%%%%%%%%%%%%%%%%%%%%%%%%%%%%%%%%%%%%%%%%%%%%%%%%%%%%%%%%%%%%%%%%%%%%%%%%%%%%%%%%%%%%%%%%%%%%%
%%
%%

\subsection{Comparison to numerical results}
\label{sec_geff_num_confirm}

We now compute the effective coupling $g_{\mathrm{eff}}^2$ numerically as defined in (\ref{eq_geff}). For the distribution function we use
\begin{equation}
 f(t,p) \simeq \frac{1}{g}\,\frac{A}{(p/B)^{\kappa_<} + (p/B)^{\kappa_>}}\,,
\label{eq_fit_distribution_lattice}
\end{equation}
which has been suggested in Ref.~\cite{Orioli:2015dxa} to approximate the distribution function at low momenta during the self-similar regime. The parameter $B$ is related to the momentum scale $K$ defined in Eq.~(\ref{eq_def_scale_K}) via $K = B\,((2-\kappa_<)/(\kappa_>-2))^{1/(\kappa_> - \kappa_<)}$, such that both are of the same order $K \sim B$ and, for $\kappa_< = 0$ and $\kappa_> = 4$, even equal $K = B$. The function in Eq.~(\ref{eq_fit_distribution_lattice}) reduces to power laws $\sim p^{-\kappa_{>}}$ for large $p \ggSim B$ and to $\sim p^{-\kappa_{<}}$ for small $p \llSim B$ momenta. 

%%
%% effective coupling schematic illustration
%%%%%%%%%%%%%%%%%%%%%%%%%%%%%%%%%%%%%%%%%%%%%%%%%%%%%%%%%%%%%%%%%%%%%%%%%%%%%%%%%%%%%%%%%%%%%%%%%%%%
\begin{figure}[t]
	\centering
	\includegraphics[scale=0.25]{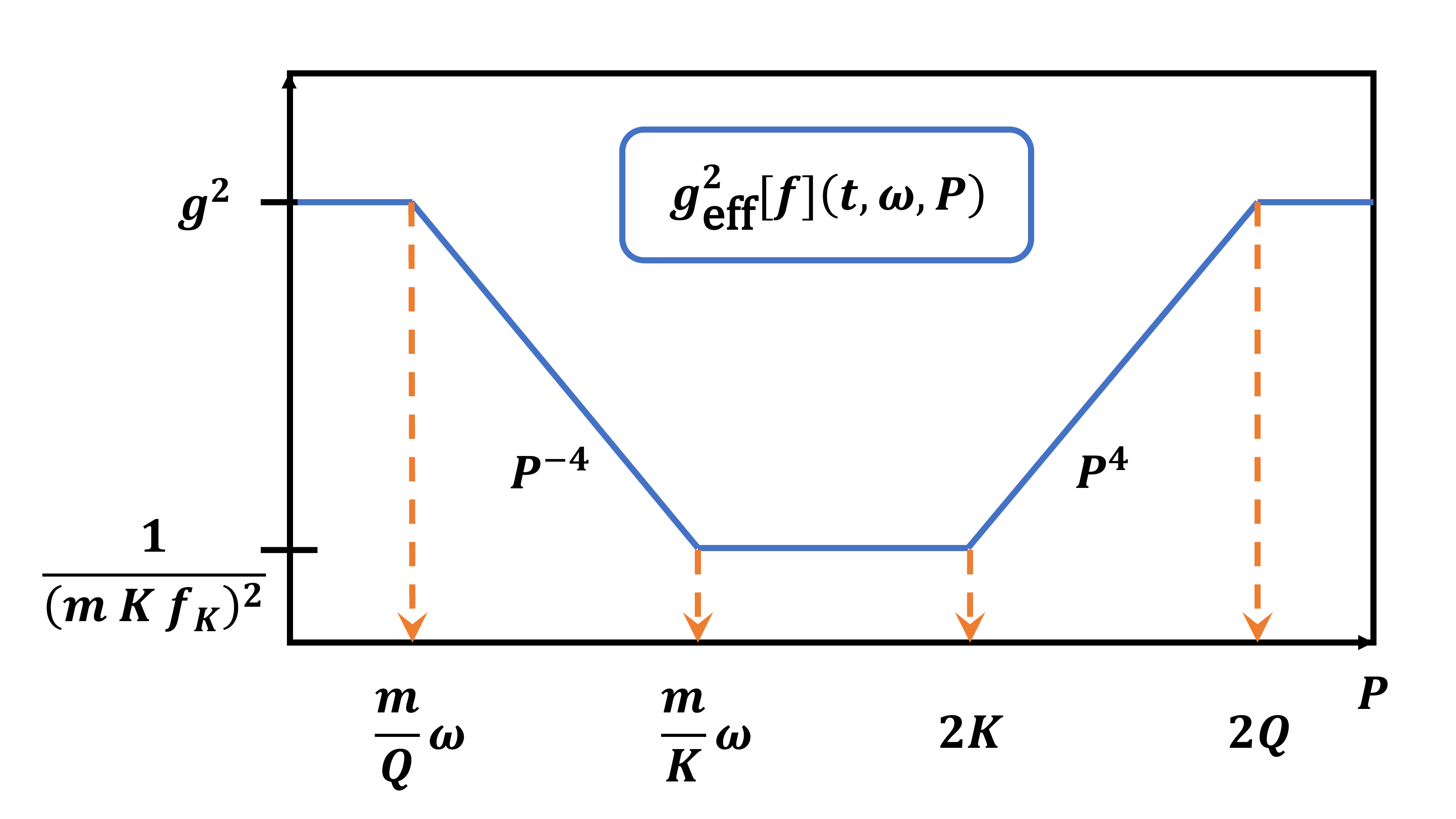}
	\caption{Schematic illustration of the effective coupling $g^2_\mathrm{eff}[f](t,\omega,\pMq)$ for fixed frequency $\omega$ that summarizes typical scales and the general functional form.}
	\label{fig_effective_coupling_sketch}
\end{figure}
%%%%%%%%%%%%%%%%%%%%%%%%%%%%%%%%%%%%%%%%%%%%%%%%%%%%%%%%%%%%%%%%%%%%%%%%%%%%%%%%%%%%%%%%%%%%%%%%%%%%
%%
%%

The effective coupling is shown in Fig.~\ref{fig_effective_coupling_omega_pp2m_and_0} for the momentum-dominated regime (labeled $\omega = 0$) and for the dispersive regime (labeled $\omega = \pMq^2/2m$) with $\kappa_< = 0$. In order to show that $g_{\mathrm{eff}}^2$ follows the parametric estimates derived in Secs.~\ref{sec_disp_regime} and \ref{sec_momentum_dominated} and is thus insensitive to the precise functional form of $f(t,p)$, we vary $3.5 \leq \kappa_> \leq 6$ and pool the resulting curves into a band for each regime. Moreover, the $1$ in the denominator of the effective coupling (\ref{eq_geff}) has been removed to be able to follow the power laws to larger momenta.\footnote{With the correct denominator the only difference would be that the scaling behavior of the effective coupling would stop and asymptotically reach the high-momentum limit $g^2$.} 
One observes that while at low momenta $g_{\mathrm{eff}}^2$ stays constant, it grows as $\pMq^4$ in the momentum-dominated regime and as $\pMq^2$ in the dispersive regime, irrespective of the detailed form of $f(t,p)$. In addition, the transition between constant and power law behavior occurs roughly at $2B$ in the prior and at $B$ in the other case. We checked that using $\kappa_< = 0.5$ leads to similar results. These observations confirm the parametric expressions for the effective coupling in Eqs.~(\ref{eq_pGGq_geff_pGGK}, \ref{eq_pGGq_geff_pLLK}, \ref{eq_pSimq_geff_pGGK}, \ref{eq_pSimq_geff_pLLK}).

Combining these scenarios with the collinear regime, we can also understand the functional form of the effective coupling for any fixed $2m\omega = p^2 - q^2$ as a function of $\pMq$. Before showing our numerical results, we use our parametric estimates to illustrate the functional form of $g_{\mathrm{eff}}^2$ and its relevant momentum scales in Fig.~\ref{fig_effective_coupling_sketch} for sufficiently low frequency differences $2m\omega \llSim (2K)^2$, where this time we also include the $1$ in the denominator of (\ref{eq_geff}). For large momenta $\pMq \ggSim 2 K$, the coupling $g_{\mathrm{eff}}^2$ is in the momentum-dominated regime and follows (\ref{eq_pSimq_geff_pGGK}), growing as a power law $\pMq^4$. In contrast, at low momenta $\pMq \llSim m\omega/K$, it is in the collinear regime and decreases as a power law $\pMq^{-4}$ due to (\ref{eq_pSimq_llW_geff}). Between these regions $m\omega/K \llSim \pMq \llSim 2K$, it takes its minimal value $g_{\mathrm{eff}}^2 \sim (m\, K\, f(t,K))^{-2}$ as in Eq.~(\ref{eq_pSimq_geff_pLLK}). Additionally, it becomes $g_{\mathrm{eff}}^2 \approx g^2$ at very large and very low momenta, parametrically given by
\begin{align}
 \label{eq_moms_for_geff_eq_g}
 \pMq \gtrsim  2\invcoh\; \quad \textrm{or} \quad \; \pMq \lesssim \frac{m}{\invcoh}\,\omega\,,
\end{align}
respectively, where we used (\ref{eq_pSimq_geff_pGGK}) and (\ref{eq_pSimq_llW_geff}). The inverse coherence scale $\invcoh$ defined in Eq.~(\ref{eq_scale_M_def}) enters the expressions because of
\begin{align}
	\label{eq_upper_scale_g_Eq_geff}
	\invcoh^2 \sim m\,g\,n \sim m\, g\, K^3\, f(t,K)\,.
\end{align}
Therefore, the momentum-dominated regime with $\omega = 0$ shown in Fig.~\ref{fig_effective_coupling_omega_pp2m_and_0} is a special case of this functional form. According to our considerations in Secs.~\ref{sec_momentum_dominated} and \ref{sec_collinear}, there is no constant region with $g_{\mathrm{eff}}^2 \sim (m\, K\, f(t,K))^{-2}$ for larger frequencies $2m\omega \ggSim (2K)^2$, but the $\pMq^{-4}$ power law changes over to $\pMq^4$ at the transition scale $\pMq^2 \approx 2m\omega$ between collinear and momentum-dominated regimes. 

%%
%% effective coupling for fixed omega 
%%%%%%%%%%%%%%%%%%%%%%%%%%%%%%%%%%%%%%%%%%%%%%%%%%%%%%%%%%%%%%%%%%%%%%%%%%%%%%%%%%%%%%%%%%%%%%%%%%%%
\begin{figure}[t]
	\centering
	\includegraphics[scale=0.23]{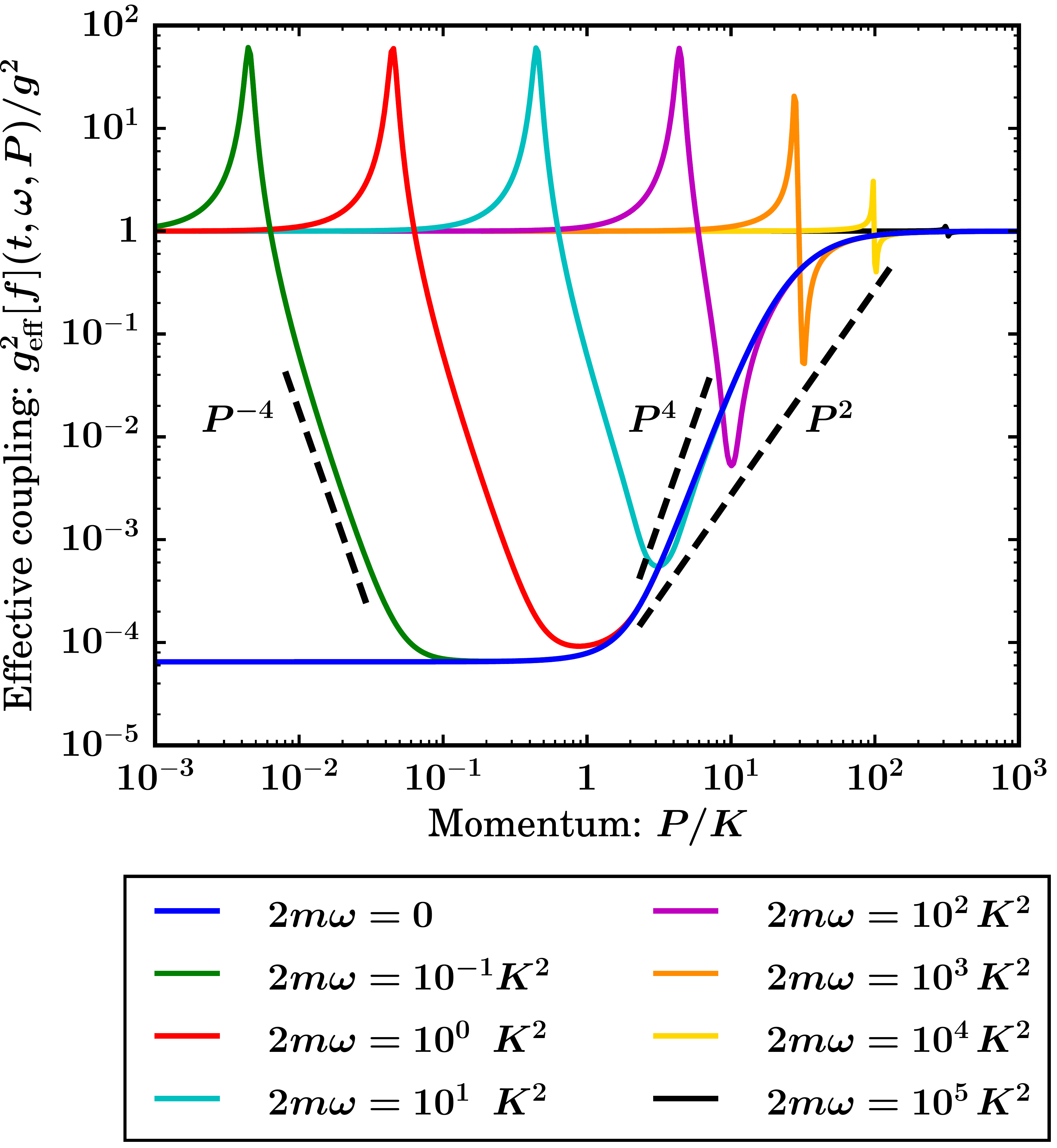}
	\caption{Effective coupling $g^2_\mathrm{eff}[f](t,\omega,\pMq)$ including the $+1$ in its denominator as a function of momentum for different values of $2m\omega$. The same distribution function and parameters are chosen as in Fig.~\ref{fig_effective_coupling_omega_pp2m_and_0} with $\kappa_> = 4$.}
	\label{fig_effective_coupling_fixed_omegas}
\end{figure}
%%%%%%%%%%%%%%%%%%%%%%%%%%%%%%%%%%%%%%%%%%%%%%%%%%%%%%%%%%%%%%%%%%%%%%%%%%%%%%%%%%%%%%%%%%%%%%%%%%%%
%%
%%

All of this is also seen in our numerical results of $g_{\mathrm{eff}}^2$ for different $2m\omega$ in Fig.~\ref{fig_effective_coupling_fixed_omegas}. For low frequency differences $2m\omega \llSim (2K)^2$ each of the parts visualized in Fig.~\ref{fig_effective_coupling_sketch} can be identified (blue, green and red curves) while for larger frequency differences $(2K)^2 \llSim 2m\omega \llSim (2\invcoh)^2$, the constant middle part is absent and the transition between the power laws occurs at $\pMq^2 \approx 2m\omega$ (cyan and violet curves), as expected from parametric estimates. At large frequency differences $2m\omega \ggSim (2\invcoh)^2$, the conditions in (\ref{eq_moms_for_geff_eq_g}) are true for almost all momenta $\pMq$ and thus, one has $g_{\mathrm{eff}}^2 \approx g^2$ such that no clear power laws are visible (yellow and black curves).\footnote{The frequency of the orange curve is of the order of $2m\omega \sim (2\invcoh)^2$, which is between the described functional forms.} 

Besides these general features, we can also understand the origin of the spikes that are visible in Fig.~\ref{fig_effective_coupling_fixed_omegas}. The upper spikes result from the negative $\textrm{Re}\,\PiR$ in the collinear regime approaching $-1$ and thus canceling the $1$ in the denominator of $g_{\mathrm{eff}}^2$. This occurs roughly at the scale when $g_{\mathrm{eff}}^2 \approx g^2$ becomes constant at low momenta. Because of $\textrm{Im}\, \PiR \llSim |\textrm{Re}\, \PiR| \approx 1$, the effective coupling becomes $g_{\mathrm{eff}}^2 \approx g^2 / (\textrm{Im}\, \PiR)^2 \gtrsim g^2$ at its maximum and exceeds there $g^2$, which results in a peak. The lower spikes are a consequence of the real part changing its sign between the collinear and momentum-dominated regimes for $\pMq \ggSim 2 K$. Then the real part vanishes $\textrm{Re}\, \PiR \approx 0$ and one again has $g_{\mathrm{eff}}^2 \approx g^2 / (\textrm{Im}\, \PiR)^2$. Since this occurs at $\pMq^2 \approx 2m\omega$, we can use the expression for the imaginary part in the dispersive regime (\ref{eq_pGGq_ImPi_pGGK}), such that the effective coupling grows as $\pMq^2$ in that case. For comparison, a corresponding power law curve is shown in Fig.~\ref{fig_effective_coupling_fixed_omegas}, which is in good accordance with the peaks of the lower spikes. 

This constitutes a detailed understanding of the functional form of $g_{\mathrm{eff}}^2$ based on parametric estimates. It also shows the importance of the inverse coherence scale $\invcoh$. All nontrivial values of the effective coupling $g_{\mathrm{eff}}^2 < g^2$ occur for momenta below this scale, such that $\invcoh$ can be regarded as the separation scale between the non-perturbative infrared region and the perturbative regime at hard momenta. 

In the language of relativistic scalar theory for large infrared occupation numbers, the inverse coherence length is related to a mass shift contained in the effective mass $M$.\footnote{The effective mass $M^2$ includes a term
\begin{align}
 \label{eq_mass_shift}
 \sim \lambda \int \mathrm{d}^3p\; \frac{f(t,\mathbf{p})}{\omega_\mathbf{p}} \,\sim\, \frac{\lambda}{m}\, n\, \sim\, \invcoh^2\,,
\end{align}
that results from local self-interactions.}
Hence, one has $M \gtrsim \invcoh$, which implies that the whole low-momentum region where the effective coupling takes nontrivial values is below the mass scale and thus, essentially non-relativistic. This has indeed been observed in lattice simulations \cite{Orioli:2015dxa,Berges:2015ixa}. Therefore, our analysis of the effective coupling should also be valid for the relativistic case.

\section{Extended range of validity of large-$N$ kinetic theory}
\label{sec_applicability}

Having understood the behavior of the effective coupling $g_{\mathrm{eff}}^2$ in the previous section, we can study its consequences for the large-$N$ kinetic theory. Different from the perturbative kinetic approach, we show here that no limitations on the typical occupancy of the distribution function exist and one can therefore consider also highly occupied systems. In particular, we demonstrate that the large-$N$ kinetic theory coincides with the standard perturbative description at large $N$ for sufficiently low typical occupancies or if only large momenta $p \gg \invcoh$ are considered. 

\subsection{Role of the effective vertex in the collision integral}
\label{sec_geff_values}

Using the results of Sec.~\ref{sec_scaling_effective_coupling}, we first discuss which values of $g_{\mathrm{eff}}^2$ actually occur in an evaluation of the collision integral (\ref{eq_collision_integral_massaged}) for incoming momentum $p$. In particular, this will allow us to identify standard perturbative descriptions in the regime where the occupancies are not large.

The integration variables relevant for the effective coupling are $q$ and $\pMq$. While the $q$-integration proceeds over all momenta, the $\pMq$ integration is limited by $\pMq_< = |p - q|$ and $\pMq_> = p + q$ and thus, not all values of $g_{\mathrm{eff}}^2$ that were discussed above occur for given $p$. The relation
\begin{align}
	\label{eq_2mw_slimits_relation}
	2m|\omega| = \pMq_< \, \pMq_>
\end{align}
provides a useful link between the frequency difference $2m\omega = p^2 - q^2$ and the integration limits.

In the dispersive regime, one has $\pMq^2 \approx 2m|\omega|$, with $\pMq \approx p$ or $\pMq \approx q$, and the resulting effective coupling is written in (\ref{eq_pGGq_geff_pGGK}), (\ref{eq_pGGq_geff_pLLK}). The $\pMq$-integration can then be simplified to 
\begin{align}
	\int_{\left|p-q\right|}^{p+q} \mathrm{d} \pMq &\, \overset{q\, \ll \, p}{\approx} \left. 2q\, \right|_{\pMq \approx p} \\ 
	\int_{\left|p-q\right|}^{p+q} \mathrm{d} \pMq &\, \overset{q\, \gg \, p}{\approx} \left. 2p\, \right|_{\pMq \approx q}\,.
\end{align}
While for $q \ll p$ the effective coupling is approximately $g_{\mathrm{eff}}^2[f]\left(t,p^2/2m,p\right)$, which stays constant during the $q$-integration, the coupling $g_{\mathrm{eff}}^2[f]\left(t,q^2/2m,q\right)$ varies for $q \gg p$.

When the momenta $q \sim p$ are of the same order, the $\pMq$-integration range becomes nontrivial. However, because of (\ref{eq_2mw_slimits_relation}) there is always a region with $\pMq^2 \approx 2m|\omega|$ as in the dispersive regime for any $p$ and $q$. The integration limits $\pMq_>^2 > 2m|\omega|$ and $\pMq_<^2 < 2m|\omega|$ are thus located around this region and correspond to the momentum-dominated and the collinear regimes, respectively. There the value of the effective coupling is given by (\ref{eq_pSimq_geff_pGGK}), (\ref{eq_pSimq_geff_pLLK}) and (\ref{eq_pSimq_llW_geff}). We note that for the special case that both momenta $p, q \llSim K$, the effective coupling takes its minimal value $g_{\mathrm{eff}}^2 \sim (m\, K\, f(t,K))^{-2}$ for all momenta $\pMq$ in the integration. 

The effective coupling becomes trivial $g_{\mathrm{eff}}^2 \approx g^2$ at sufficiently large frequency differences $2m|\omega| \ggSim (2\invcoh)^2$. Because of $2m\omega = p^2 - q^2$, this statement depends on the integration variable $q$ in the collision integral and may occur as part of the integration. Similarly, this happens at low and high momentum differences $\pMq$ as in (\ref{eq_moms_for_geff_eq_g}). 

In this context, an interesting case is when the incoming momentum exceeds the inverse coherence length $p \gg \invcoh$. Then the effective coupling stays mostly trivial, i.e., $g_{\mathrm{eff}}^2 \approx g^2$ during the $q$-integration. For $p \gg q$ and $p \ll q$ this results from $\pMq \gg \invcoh$. The $q$-momentum range where $g_{\mathrm{eff}}^2$ takes nontrivial values is located around $p \sim q$ and, because of $\pMq \sim \invcoh \ll q$, it is small when compared to typical $q$ values and to the whole $q$-integration range that proceeds from $0$ to $\infty$. This shows that the large-$N$ kinetic theory reduces to the perturbative kinetic theory if dynamics at only large momenta $p \gg \invcoh$ is studied.  

Furthermore, $g_{\mathrm{eff}}^2$ also becomes trivial when $K \gtrsim \invcoh$, which is independent of integration parameters. This condition is equivalent to $m\, g\, K\, f(t,K) \lesssim 1$, which, as will be shown in Eq.~(\ref{eq_interference_pert}), corresponds to $\zeta\, f(t,K) \lesssim 1$, where $\zeta$ is the diluteness parameter. Hence, when typical occupation numbers are sufficiently small, the effective coupling becomes $g^2$ for all values of the integration parameters and the large-$N$ kinetic theory reduces to the perturbative kinetic theory.

\subsection{Interference terms and their large-$N$ resummation}
\label{sec_interference}

\begin{figure}[t]
	\unitlength=1mm
	\centering
	\includegraphics[scale=0.95]{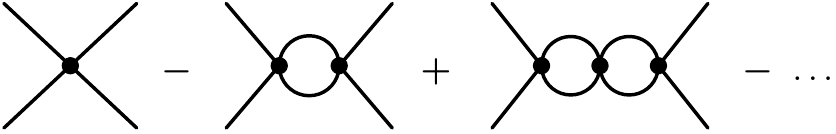}
	\caption{Perturbative expansion of the effective coupling $g_{\mathrm{eff}}^2$.}
	\label{fig_2to2_scatterings}
\end{figure}

The collision integrals of perturbative and large-$N$ kinetic theories differ in the coupling, which is constant $g^2$ or resummed $g^2_{\mathrm{eff}}$, respectively. To understand the range of validity of the theories due to interference, we consider a series of diagrams that all contribute to the elastic $2 \leftrightarrow 2$ channel, as depicted in Fig.~\ref{fig_2to2_scatterings}. Since the retarded self-energy $\PiR$ as defined in Eq.~(\ref{eq_1loop_ret_self_energy_massaged}) is originally a convolution of the distribution function and the retarded propagator, it corresponds to a ``blob'' connecting two vertices. With the results of Sec.~\ref{sec_scaling_effective_coupling}, it can be parametrically estimated as
\begin{equation}
 \PiR\left(t,\omega,\pMq\right) \sim m\, g\, K\, f(t,K) \sim \left(\zeta\, f(t,K)\right)^{2/3},
 \label{SKT_one_loop_retarded_self_energy}
\end{equation}
where we have included the diluteness parameter $\zeta \sim m\, g\, \invcoh$ with the scale $\invcoh^2 \sim m\, g\, K^3\, f(t,K)$, which are defined in Eqs.~(\ref{eq_diluteness_param_def}) and (\ref{eq_scale_M_def}). As long as $|\PiR| \ll 1$, the series of chain diagrams in Fig.~\ref{fig_2to2_scatterings} corresponds to a (perturbative) loop expansion whose leading term in the collision integral is given by the coupling constant $g^2$. Hence, 
\begin{align}
 \label{eq_interference_pert}
 m\,g\,K\,f(t,K)\, \overset{\mathrm{pert}}{\ll}\, 1 \quad \Leftrightarrow \quad \zeta\, f(t,K) \, \overset{\mathrm{pert}}{\ll}\, 1
\end{align}
is a necessary condition for the validity of the perturbative kinetic theory.\footnote{Although we consider a non-relativistic theory with a quadratic dispersion relation, it is enlightening to check how this condition translates into the language of a relativistic scalar theory. Assuming that the effective mass $M$ is of the order of $\invcoh$, one can identify $\lambda \sim \zeta$. Hence, perturbative kinetic theory is valid for small enough occupancies up to the inverse coupling $f(t,K) \ll 1/\lambda$, which is consistent with what is commonly known (see, e.g., \cite{Mueller:2002gd,Berges:2015kfa}.}
The equivalent condition on the right hand side restricts the occupation numbers for that case. Otherwise, when $|\PiR| \gtrsim 1$, all of the considered scattering processes are of the same order and interfere, which leads to the break-down of the perturbative approach.

This problem is absent in the large-$N$ kinetic description because of the effective coupling $g^2_{\mathrm{eff}}$ in \eqref{eq_geff}. Its perturbative expansion corresponds to the series of scattering processes that was depicted in Fig.~\ref{fig_2to2_scatterings}. Hence, the effective coupling appears as a geometric series at small $\PiR$ where the series converges. More precisely, it resums all diagrams emerging at NLO in a $1/N$ expansion~\cite{Aarts:2002dj,Berges:2008wm}. 

Hence, the interference of an infinite number of scattering processes is encoded into the vertex resummation in the large-$N$ kinetic theory and results in an effective coupling $g^2_{\mathrm{eff}}$. In this effective kinetic theory, characteristic occupancies are not restricted. Thus, it comprises and extends the range of validity of standard perturbative kinetic descriptions.

\subsection{Cross section and mean free path}
\label{sec_cross_section}

In Sec.~\ref{sec:largeN}, we have argued that the large-$N$ kinetic theory appears as a consistent quasiparticle description in the large-$N$ limit since the spectral function can be approximated by an on-shell form in the collision integral. Moreover, this effective theory extends perturbative kinetic approaches and can also be applied to highly occupied systems, as we argued in this section. In addition to these general arguments, we give here a more illustrative explanation of the validity of the large-$N$ kinetic theory in terms of (perturbative) concepts as cross sections and mean free paths. 

We first get an expression for the cross section $\sigma$ for elastic scattering in the presence of the vertex corrections encoded in the effective coupling. We start with a general Lorentz invariant expression of the differential cross section for $2 \leftrightarrow 2$ scatterings in the center-of-mass system (CMS) (see, e.g.,~\cite{Halzen:1984mc})
\begin{align}
\label{eq_cross_section_diff}
\frac{\mathrm{d} \sigma}{\mathrm{d} \Omega} = \frac{|\mathcal{M}|^2}{64 \pi^2 s}\,.
\end{align}
In the considered non-relativistic limit, the Mandelstam variable $s$ simplifies to $s \approx 4 m^2$. The matrix element for the perturbative kinetic theory can be expressed as $|\mathcal{M}|^2 \sim m^4 g^2/N$, where we included the number of field components $N$ explicitly. For the large-$N$ kinetic theory, the coupling constant is replaced by the effective coupling $g^2 \mapsto g_{\mathrm{eff}}^2[f]\left(t, \omega, \pMq \right)$. In the CMS, in- and outgoing momenta are equal $p = q$ and thus, one may set $\omega = 0$. 

The total cross section can be obtained from (\ref{eq_cross_section_diff}) by integration. For this, we substitute the integration variables from angular variables to $\pMq^2 = 2p^2 - 2p^2 \cos \theta_{\mathbf{pq}}$ with $\mathrm{d} \Omega = 2\pi \, \mathrm{d}\!\cos \theta_{\mathbf{pq}} = -(\pi / p^2)\; \mathrm{d}\pMq^2$, and arrive at
\begin{align}
\label{eq_cross_section_tot}
\sigma = \frac{1}{64 \pi\, (2p)^2\,m^2} \int_{0}^{(2p)^2}\!\!\!\mathrm{d}\pMq^2 \; |\mathcal{M}|^2\left(t, \omega = 0, \pMq \right)\,.
\end{align}

Hence, for constant coupling $g^2$ as in the standard (perturbative) kinetic theory, one has the familiar expression\footnote{The $1/N$ factor is often omitted in such expressions \cite{Arnold:2007pg} but in our case, it is important to be consistent with large-$N$ arguments. Also recall that we consider here the non-relativistic limit. For an (ultra-)relativistic theory, one would have $s \approx 4 p^2$, with $p = |\mathbf{p}|$, which would lead to $\sigma_{\mathrm{pert}}^{\mathrm{rel}} \sim \lambda^2/(p^2N)$. In equilibrium, typical momenta are of the order of the temperature scale $p \sim T$, which would lead to familiar expressions for the cross section and the mean free path \cite{Arnold:2007pg}.}
\begin{align}
\sigma_{\mathrm{pert}} \sim \frac{m^2 g^2}{N} \propto \frac{\lambda^2}{m^2\,N}\,,
\end{align}
where we included the relativistic coupling $\lambda$ for comparison. The same expression $\sigma_{\mathrm{eff}} \approx \sigma_{\mathrm{pert}}$ is found for the effective coupling whenever it reduces to $g^2$, which occurs at sufficiently low typical occupancies $\zeta\, f(t,K) \lesssim 1$ or at large momenta $p \gg \invcoh$ (Secs.~\ref{sec_geff_values} and \ref{sec_geff_num_confirm}). Otherwise $g_{\mathrm{eff}}^2[f]\left(t, \omega = 0, \pMq \right)$ is in the momentum-dominated regime that was discussed in Sec.~\ref{sec_momentum_dominated}. One then has
\begin{eqnarray}
\sigma_{\mathrm{eff}} &\overset{\invcoh \, \ggSim p \, \ggSim \, K}{\sim}& \frac{1}{N(K\, f(t,K))^2} \; \frac{p^4}{K^4}  \\
\sigma_{\mathrm{eff}} &\overset{p \, \sim \, K}{\sim}& \frac{1}{N(K\, f(t,K))^2}\,.
\end{eqnarray}

Let us now proceed with the mean free path $L$ of quasiparticles between two collisions for typical momenta $K$. A condition for a valid quasiparticle picture is that the mean free path $L$ of quasiparticles is larger than their spatial extent, which can be estimated by their de Broglie wavelength $\lambda_{\textrm{DB}} \sim 1/K$~\cite{Arnold:2007pg}, such that
\begin{align}
\label{eq_LGGlambda}
L \gg \frac{1}{K}\,.
\end{align}
The mean free path of classical particles is usually estimated by means of the cross section and of the particle number density as $L_{\mathrm{class}} \sim 1/(\sigma n)$. Since the frequency of collisions that a typical particle encounters may be enhanced by the Bose factor $(1 + f(t,K))$ \cite{Baier:2000sb}, an alternative estimate of the mean free path is
\begin{align}
\label{eq_L_Bose_def}
 L \sim \frac{1}{\sigma \, n \, (1 + f(t,K))}\,.
\end{align}
Because of $L_{\mathrm{class}} \geq L$, the quasiparticle relation (\ref{eq_LGGlambda}) is valid for $L_{\mathrm{class}}$ whenever this is the case for $L$. Therefore, we will use $L$ in Eq.~(\ref{eq_L_Bose_def}) for our estimates. 

On general grounds, we can already compare the large-$N$ effective mean free path (referred to as $L_{\mathrm{eff}}$) and the one resulting from standard (perturbative) kinetic theory $L_{\mathrm{pert}}$. Because of $g_{\mathrm{eff}}^2 \leq g^2$, one has 
\begin{align}
\sigma_{\mathrm{eff}} \leq \sigma_{\mathrm{pert}} \quad \Leftrightarrow \quad L_{\mathrm{eff}} \geq L_{\mathrm{pert}}\,.
\end{align}
Hence, the effective coupling leads to a reduced cross section and an increased mean free path, counteracting in this way even high quasiparticle densities. 

To be more specific, for the perturbative case of constant $g^2$ interactions, the mean free path can be estimated as
\begin{align}
\label{eq_Lmfp_pert}
L_{\mathrm{pert}} \sim \frac{N}{K\, (m\,g\,K\,f(t,K))^2}\,.
\end{align}
The same expression is encountered for the large-$N$ kinetic theory in the cases when the effective coupling $g_{\mathrm{eff}}^2$ reduces to $g^2$. Otherwise, one has $\zeta\, f(t,K) \gtrsim 1$ and
\begin{align}
L_{\mathrm{eff}} \overset{p\, \sim\, K}{\sim} \frac{N}{K}\,.
\label{eq_Lmfp_eff_pLLK}
\end{align}
For perturbative kinetic theory, the mean free path (\ref{eq_Lmfp_pert}) and the imposed restriction on occupation numbers $m\,g\,K\,f(t,K)\, \ll \, 1$ in (\ref{eq_interference_pert}) imply $L_{\mathrm{pert}} \gg 1/K$ and condition (\ref{eq_LGGlambda}) is satisfied. 

In the case of the large-$N$ kinetic theory, no restrictions on the occupation numbers have been posed. Whenever the effective coupling reduces to the coupling constant at momenta of the order $K$ at vanishing $\omega$, one has $m\,g\,K\,f(t,K)\, \lesssim \, 1$ and for large $N$, this leads to $L_{\mathrm{pert}} \gtrsim N/K \gg 1/K$. Otherwise, we can use (\ref{eq_Lmfp_eff_pLLK}) with $L_{\mathrm{eff}} \sim N/K \gg 1/K$, which is again consistent with the underlying assumption of large $N$ and satisfies the condition (\ref{eq_LGGlambda}). 

In summary, based on the interference argument in Sec.~\ref{sec_interference}, typical occupation numbers are restricted to $\zeta\,f(t,K) \ll 1$ for perturbative kinetic theory. Since $f(t,K)$ is a dynamical variable, this condition is time dependent. On the other hand, no restrictions on occupation numbers exist for large-$N$ kinetic theory. In addition, Eq.~(\ref{eq_LGGlambda}) is satisfied in both cases, which contributes to a consistent quasiparticle interpretation.

%%%%%%%%%%%%%%%%%%%%%%%%%%%%%%%%%%%%%%%%%%%%%%%%%%%%%%%%%%%%%%%%%%%%%%%%%%%%%%%%%%%%%%%%%%%%%%%%%%%%
%%%%%%%%%%%%		SECTION: NUMERICAL SOLUTION
%%%%%%%%%%%%%%%%%%%%%%%%%%%%%%%%%%%%%%%%%%%%%%%%%%%%%%%%%%%%%%%%%%%%%%%%%%%%%%%%%%%%%%%%%%%%%%%%%%%%

\section{Nonthermal fixed point from large-$N$ kinetic theory}

\label{sec_solution}

In this section, we determine the distribution function in the highly occupied self-similar regime at low momenta of scalar systems \cite{Orioli:2015dxa}. For the first time, we compute the universal scaling function $\fs(p)$ within the large-$N$ kinetic theory, performed in three spatial dimensions. This is done numerically by solving a rescaled version of the kinetic equation that leads to the fixed point solution $\fs(p)$. We present our numerical approach and discuss the solution. Although we focus here on the self-similar regime, our numerical approach can be applied to more general situations of the thermalization process of scalar systems. 

If not stated otherwise, all dimensionful quantities in this section are provided in suitable powers of $2m$, and occupation numbers are rescaled by $4m^2 g$. Hence, time, momenta and occupation numbers are obtained by the rescalings $t \mapsto 2m\,t$, $p \mapsto p / (2m)$ and $f \mapsto f / (4m^2 g)$.

\subsection{Review of lattice simulation results}
\label{sec_lattice_results}

It was found in classical-statistical lattice simulations that non-relativistic and relativistic $O(N)$ symmetric scalar field theories for different number of components $N$ approach a universal scaling regime at low momenta where the distribution function is large $\zeta\, f(t,K) \gtrsim 1$ and follows a self-similar evolution~\cite{Orioli:2015dxa,Moore:2015adu,Schachner:2016frd}
\begin{align}
 \label{eq_selfsim}
 f(t,p) = \ttref^\alpha \fs \left(\ttref^\beta\, p\right).
\end{align}
The scaling exponents $\alpha$ and $\beta$ as well as the scaling function $\fs(p)$ are the same for a wide range of initial conditions once the evolution is sufficiently close to its nonthermal fixed point \cite{Berges:2008wm}. Since these properties are universal in the sense that they are approximately the same even in different theories,  they form a far-from-equilibrium universality class \cite{Berges:2014bba,Orioli:2015dxa}. In three spatial dimensions, the scaling exponents have been measured to be close to the values
\begin{align}
 \label{eq_scaling_exp_ab}
 \alpha = \frac{3}{2} ~, \qquad \beta = \frac{1}{2}\,.
\end{align}
The scaling function $\fs(p)$ was observed to be approximately described by the functional form (\ref{eq_fit_distribution_lattice}). This form can be approximated by two power laws for large and small momenta around a typical (stationary) infrared scale $\Ks$ being the momentum where the particle number density is dominated according to (\ref{eq_def_scale_K})
\begin{align}
\label{eq_spectrum_lattice}
 \fs(p) \overset{p \ggSim \Ks}{\sim} p^{-\kappa_>}, \quad \fs(p) \overset{p \llSim \Ks}{\sim} p^{-\kappa_<}.
\end{align}
The lower exponent has been observed to lie within the range $0 \leq \kappa_< \leq 1/2$ \cite{Orioli:2015dxa,Moore:2015adu}. The larger spectral exponent $\kappa_>$ has been investigated in different studies and has often been related to strong wave turbulence. Considering different systems in $d = 3$ spatial dimensions, exponents were found for a range of different phenomena, ranging between $4$ and $5$ \cite{Berges:2008wm,Scheppach:2009wu,Berges:2010ez,Gasenzer:2011by,Nowak:2011sk,Mathey:2014xxa,Orioli:2015dxa,Moore:2015adu}. In some studies \cite{Nowak:2011sk,Gasenzer:2011by,Mathey:2014xxa,Karl:2013kua}, the observed exponent could be understood as a superposition of two different power laws, associated to different physical origins, or attributed to finite size effects. 
Simulations within the 2PI framework confirm the existence of a nonthermal fixed point in the infrared with properties (\ref{eq_spectrum_lattice}) and show that it even survives at moderate couplings \cite{Berges:2008wm,Berges:2016nru}. 

With the observed values for the scaling exponents $\alpha$ and $\beta$ in (\ref{eq_scaling_exp_ab}), the dynamics can be understood as an inverse particle cascade that occupies low momenta towards the formation of a Bose-Einstein condensate far from equilibrium \cite{Berges:2012us}. Since the condensate formation time grows with volume $t_{\mathrm{cond}} \sim V^{1/\alpha}$ while the growth of the zero momentum mode follows a power law behavior in time $f(t,\mathbf{0}) \sim t^\alpha$ \cite{Nowak:2012gd,Orioli:2015dxa,Schachner:2016frd}, for an infinitely large system, no condensate is formed.

\subsection{Self-similar solution}

Because of its large occupation numbers $\zeta\, f(t,K) \gtrsim 1$, this scaling region cannot be described by perturbative kinetic theory, the range of validity of which is restricted by Eq.~(\ref{eq_interference_pert}). Instead, the large-$N$ kinetic theory can be applied. Indeed, it was shown in Ref.~\cite{Orioli:2015dxa} that it correctly reproduces the observed scaling exponents (\ref{eq_scaling_exp_ab}). In the following, we extend this calculation by also obtaining the scaling function $\fs$, and thus, we provide a complete scaling solution of the large-$N$ kinetic theory in the low-momentum region. 

To find the scaling form of the distribution function in the infrared, we first follow Refs.~\cite{Micha:2004bv,Orioli:2015dxa} and plug the self-similarity ansatz (\ref{eq_selfsim}) into the effective kinetic equation (\ref{eq_effective_Boltzmann_equation}). With the rescaled momentum $\tilde{p} = \ttref^\beta\, p$, the left hand side can be expressed as
\begin{align}
	\frac{\partial}{\partial t} f(t,p)
	= \ttref^{\alpha - 1} \left[ \alpha \fs(\tilde{p} ) 
	+ \beta\, \tilde{p} \, \frac{\partial \fs(\tilde{p})}{\partial \tilde{p}} \right].
\end{align}
Similarly, one can rescale the collision integral\footnote{\label{foot_omit_1_in_geff}Since we are interested in the highly occupied infrared region, we assume $f \gg 1$ and $\zeta\, f(t,K) \gg 1$, neglecting the $1$ in the denominator of the effective coupling $g_{\mathrm{eff}}^2$ in (\ref{eq_geff}), which enables an exact scaling solution of the collision integral. We have checked that including $1$ only changes the distribution at large momenta outside the scaling region where $f \lesssim 1/\zeta$, which can be surpassed on the finite momentum grid by taking a sufficiently large occupation number $f(t,K)$.}
\begin{equation}
	C[f](t,p) = \ttref^{\mu}\, C\left[\fs\right]\left(\tilde{p}\right).
\end{equation}
Using $\omega_p = p^2/2m$, one finds with the expressions for the collision integral (\ref{eq_collision_integral_raw}), (\ref{eq_geff}) and (\ref{eq_1loop_ret_self_energy_raw})
\begin{align}
 \PiR\left( t, \omega_p - \omega_q, \pMq \right) &= \ttref^{\alpha - \beta} \,\PiR\left(\omega_{\tilde{p}} - \omega_{\tilde{q}}, \tilde{\pMq} \right), \\
 g_{\mathrm{eff}}^2[f]\left( t, \omega_p - \omega_q, \pMq \right) &\approx \ttref^{-2(\alpha - \beta)} \,g_{\mathrm{eff}}^2[f]\left(\omega_{\tilde{p}} - \omega_{\tilde{q}}, \tilde{\pMq} \right),
\end{align}
which leads to the scaling exponent of the collision integral \cite{Orioli:2015dxa}
\begin{align}
 \mu = \alpha - 2\beta\,.
\end{align}
Setting $\alpha - 1 = \mu$, we get rid of the explicit time factors on both sides of the kinetic equation and we arrive at $\beta = 1/2$. This value of $\beta$ is even the same in different spatial dimensions $d$ \cite{Orioli:2015dxa}. Moreover, particle number density conservation leads to the additional relation $\alpha = 3 \beta$ for the considered case of $d = 3$. Hence, the computed values agree with the observations (\ref{eq_scaling_exp_ab}).

%%
%%%%%%%%%%%%%%%%%%%%%%%%%%%%%%%%%%%%%%%%%%%%%%%%%%%%%%%%%%%%%%%%%%%%%%%%%%%%%%%%%%%%%%%%%%%%%%%%%%%%
\begin{figure}[t]
	\centering
	\includegraphics[scale=0.23]{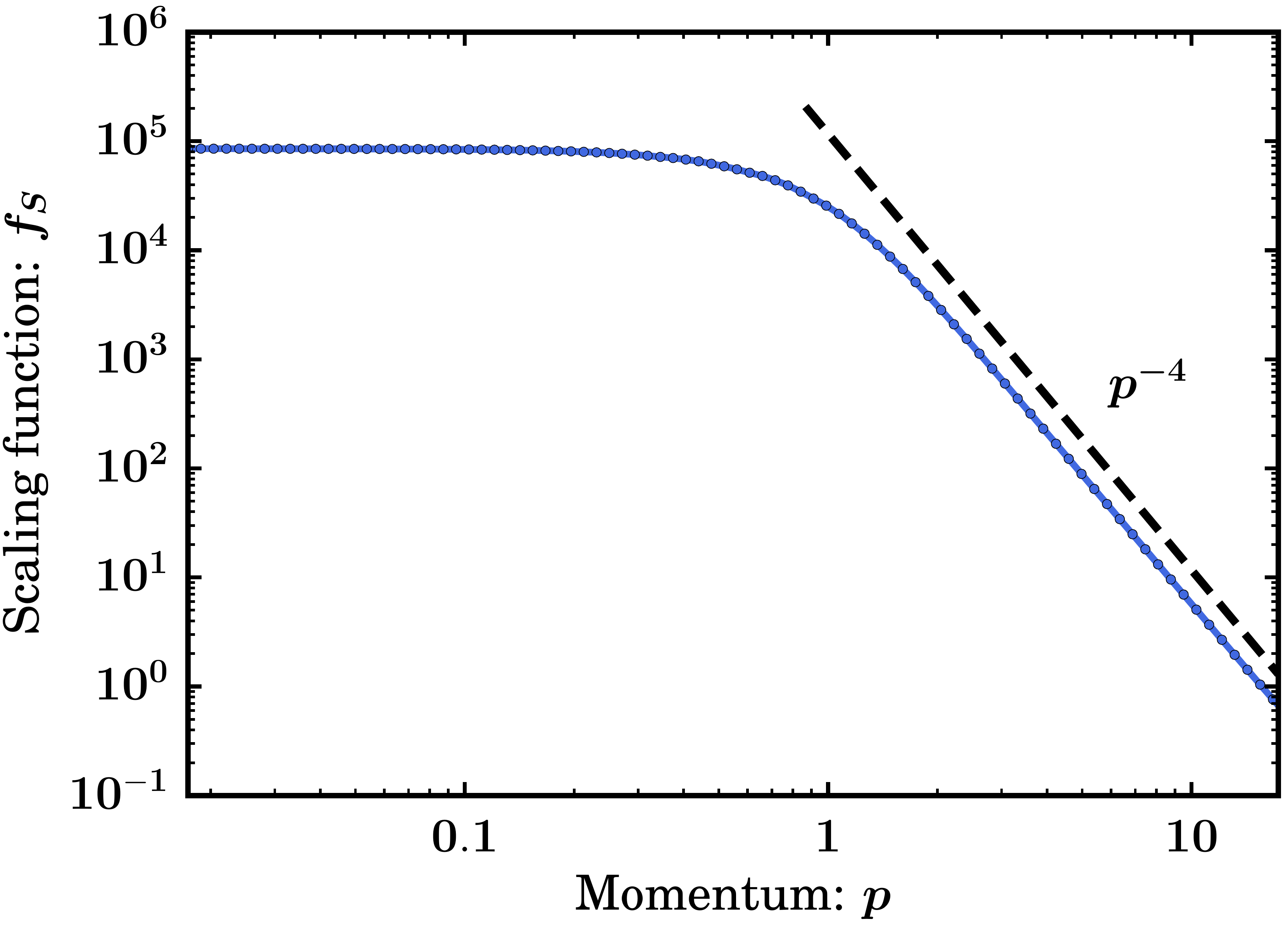}
	\caption{Scaling function $\fs(p)$ computed by solving (\ref{eq_fixed_point_kineq}) numerically. The error bars (see Sec.~\ref{sec_numerics_of_LNKT}) are smaller than the line width.}
	\label{fig_alpha15_beta05_sol}
	%	\end{minipage}
\end{figure}
%%%%%%%%%%%%%%%%%%%%%%%%%%%%%%%%%%%%%%%%%%%%%%%%%%%%%%%%%%%%%%%%%%%%%%%%%%%%%%%%%%%%%%%%%%%%%%%%%%%%
%%
%%

\begin{figure}[t]
	\centering
	\includegraphics[scale=0.23]{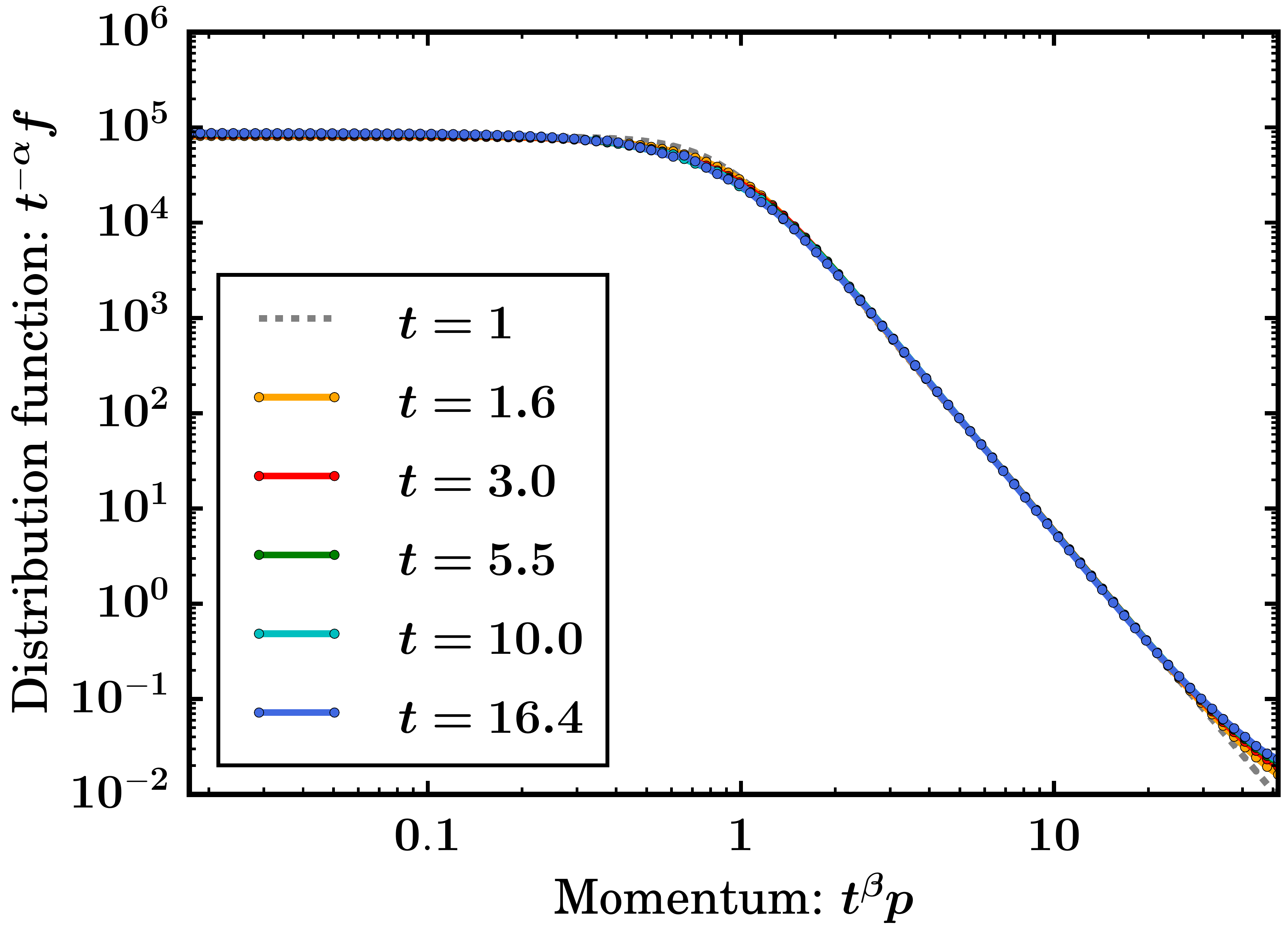}
	\caption{Rescaled distribution function $\tilde{f} = \ttref^{-\alpha} f$ as a function of rescaled momentum $\tilde{p} = \ttref^{\beta} p$ at different times. The system is initialized at $t = 1$ with the distribution function given by (\ref{eq_fit_distribution_lattice}) (gray dashed line) with parameters $A = 80000$, $B = 0.867$, $\kappa_< = 0$ and $\kappa_> = 3.9$. The rescaled distribution quickly approaches its stationary form $\fs$.}
	\label{fig_alpha15_beta05_final_evol}
\end{figure}

The remaining fixed point equation then reads 
\begin{equation}
	\frac{3}{2}\, \fs(\tilde{p}) + \frac{1}{2}\,\tilde{p} \, \frac{\partial \fs(\tilde{p})}{\partial \tilde{p}} = C[\fs] (\tilde{p}).
	\label{eq_fixed_point_kineq}
\end{equation}
We solve this equation numerically for the scaling function $\fs(\tilde{p})$. While details of our numerical approach are postponed to the following subsections, we present here its solution in Fig.~\ref{fig_alpha15_beta05_sol}. The scaling function has the following properties:
\begin{align}
 \label{eq_fS_form_proporties}
 \fs(\tilde{p}) \overset{\tilde{p} \llSim \Ks}{\sim} \text{const}, \quad \fs(\tilde{p}) \overset{\tilde{p} \ggSim \Ks}{\sim} \tilde{p}^{-4}\,,
\end{align}
with the transition (``bending'') scale at $\Ks = 1$, where we used its definition in Eq.~(\ref{eq_def_scale_K}). The exponents $\kappa_i$ of the distribution at larger and lower momenta compare well with former studies (see Sec.~\ref{sec_lattice_results}). The functional form of the here obtained $\fs$ is so similar to Eq.~(\ref{eq_fit_distribution_lattice}), which was used in Ref.~\cite{Orioli:2015dxa} to approximate $f_S$ on the lattice, that it is difficult to visually distinguish between them, as will be clearer in Fig.~\ref{fig_alpha15_beta05_final_evol}. We yet note that there are small deviations from that simple form in the bending region at the scale $\Ks$. 

Apart from the scaling form, we can also compare non-universal quantities that depend on parameters of our system like mass or particle number density. These are the amplitude $\fs(\Ks)$ and the typical momentum $\Ks$. To be more specific, we compare these quantities to the classical-statistical lattice results from non-relativistic scalar theory shown in Fig.~3 of Ref.~\cite{Orioli:2015dxa}. There, the rescaled distribution function within the self-similar regime $\overline{\fs} = \left(t/\trefTwo\right)^{-\alpha} f$ is plotted as a function of the rescaled momentum $\overline{p} = \left(t/\trefTwo\right)^{\beta} p$, where we used the overline notation to distinguish from our conventions for the scaling function and rescaled momentum. The two notations are related by
\begin{align}
 \overline{\fs} \left(\overline{p}\right) = \trefTwo^\alpha\, \fs \left(\tilde{p}\right), \quad \overline{p} = \trefTwo^{-\beta} \, \tilde{p} .
\end{align}
The reference time was set to $\trefTwo = 300$ in the same units as used here. 
With this, we can transform the typical momentum $\Ks$ and the amplitude $\fs(\Ks)$ obtained here within large-$N$ kinetic theory to the overline notation, as 
\begin{align}
\label{eq_Ks_diff_notations}
 \overline{\Ks} = \trefTwo^{-\beta}\, \Ks \approx 0.0577,
\end{align}
and $\overline{\fs}(\overline{\Ks}) = \trefTwo^{\alpha} \fs(\Ks)$, and compare them to the corresponding quantities in Ref.~\cite{Orioli:2015dxa}. Since in our prescription of the large-$N$ kinetic theory for very high occupation numbers (see footnote~\ref{foot_omit_1_in_geff}) the amplitude drops out of the kinetic equation, it can thus be adjusted arbitrarily even after the simulation, depending on the particle number density in the system $n \sim \fs(\Ks)\, \Ks^3 = \overline{\fs}(\overline{\Ks})\,\overline{\Ks}^3$. Hence, it is not suitable for a comparison with the lattice results. On the other hand, being independent of $n$, the scale $\Ks$ (or $\overline{\Ks}$) is fixed by the mass parameter $m$ (or by both $m$  and the reference time $\trefTwo$) and enables a quantitative comparison between large-$N$ and lattice simulation results. In Fig.~3 of Ref.~\cite{Orioli:2015dxa} the transition scale $\overline{\Ks}$ is located within the momentum range $0.05 \lesssim \overline{\Ks} \lesssim 0.08$. This is consistent with our result from the large-$N$ kinetic theory (\ref{eq_Ks_diff_notations}). 

We note that the small deviations between large-$N$ kinetic and lattice results may have different reasons. First of all, we use the large-$N$ kinetic theory at NLO and omit higher orders in $1/N$. Moreover, the functional form measured on the lattice in Ref.~\cite{Orioli:2015dxa} may suffer from finite-time effects and may slightly change at later times beyond the simulation times shown there.\footnote{Similar finite-time artifacts have been observed for non-Abelian gauge theory, where a kinetic description compared well with lattice results but showed slight differences at late times beyond the time of lattice simulations when being closer to the nonthermal fixed point \cite{York:2014wja}.} 
And finally, regarding the discussion below Eq.~(\ref{eq_spectrum_lattice}), the observed power law on the lattice could be a superposition of power laws with different origins. Therefore, it was of great importance to pin down the power law exponent $4$ in Eq.~(\ref{eq_fS_form_proporties}) that can be associated with the large-$N$ kinetic theory\footnote{See also Ref.~\cite{Chantesana:2018qsb} for an alternative approach to this problem and an outcome consistent with our results.} to be able to distinguish it from possibly other contributions. 

We have seen that the large-$N$ kinetic theory provides an even quantitatively good description of classical-statistical lattice data. This confirms its applicability to systems with very high occupation numbers, extending perturbative kinetic frameworks. The scaling function in Fig.~\ref{fig_alpha15_beta05_sol} and its properties are the main results of this section.

\subsection{Numerical setup}
\label{sec_numerical_setup}

Here we discuss the numerical setup that led to the scaling solution in Fig.~\ref{fig_alpha15_beta05_sol}. In order to solve the fixed point equation (\ref{eq_fixed_point_kineq}), we start again with the full kinetic equation in (\ref{eq_effective_Boltzmann_equation}) with the collision integral $C[f]$ as given by (\ref{eq_collision_integral_massaged}). Our strategy is to rescale the kinetic equation such that it relaxes to the fixed point equation with time. With this, we follow Ref.~\cite{York:2014wja} where this strategy was used for the self-similar region at hard momenta in non-Abelian gauge theory. Therefore, instead of using a self-similarity ansatz, we rescale the distribution function and momenta according to
\begin{equation}
	f(t,p) \equiv \ttref^\alpha \tilde{f} \left(t,\ttref^\beta\, p\right) \equiv \ttref^\alpha \tilde{f} \left(t,\tilde{p}\right),
	\label{NUM_sclaing_relation_distribution}
\end{equation}
with the values for the scaling exponents from (\ref{eq_scaling_exp_ab}). Comparing (\ref{NUM_sclaing_relation_distribution}) to the self-similar evolution (\ref{eq_selfsim}), one finds that the scaling function is the stationary limit of this rescaled distribution $\tilde{f} \left(t,\tilde{p}\right) \rightarrow \fs \left(\tilde{p}\right)$. 

Plugging (\ref{NUM_sclaing_relation_distribution}) into the kinetic equation (\ref{eq_effective_Boltzmann_equation}), one arrives at the rescaled kinetic equation 
\begin{equation}
	\frac{\partial \tilde{f}(t,\tilde{p})}{\partial \log t}
	= - \left[
	\alpha \tilde{f}(t,\tilde{p}) + \beta \tilde{p} \, \frac{\partial \tilde{f}(t,\tilde{p})}{\partial \tilde{p}} - C[\tilde{f}] (t,\tilde{p}) 
	\right],
	\label{NUM_rescaled_kinetic_equation}
\end{equation}
since the explicit time factors cancel. Note that Eq.~(\ref{NUM_rescaled_kinetic_equation}) is equivalent to the original kinetic equation (\ref{eq_effective_Boltzmann_equation}) but reduces to the fixed point form (\ref{eq_fixed_point_kineq}) when $\tilde{f}$ becomes time independent. Hence, a stationary solution of this equation corresponds to the scaling function $\fs(\tilde{p})$, as has been noted above. In this sense, and because of the logarithmic time derivative $\partial \tilde{f} / \partial \log t$, it can be regarded as a relaxation algorithm for $\fs(\tilde{p})$ in time. 

Moreover, the overall amplitude of $\tilde{f}$ drops out of the kinetic equation (\ref{NUM_rescaled_kinetic_equation}) because we neglect the $1$ in the denominator of the effective coupling (\ref{eq_geff}) (see also footnote (\ref{foot_omit_1_in_geff})). This corresponds to the high-occupancy limit $\tilde{f}(t,\tilde{K}) \rightarrow \infty$ with $\tilde{f}(t,\tilde{p})/\tilde{f}(t,\tilde{K})$ kept fixed for each momentum. Here $\tilde{K}$ is the typical (rescaled) infrared momentum scale where particle number density $n$ is dominated with respect to the distribution $\tilde{f}$. Similarly, the coupling constant $g$ also drops out of the kinetic equation. 

To numerically solve (\ref{NUM_rescaled_kinetic_equation}), we discretize time logarithmically with constant $\Delta\!\log t = \log t_{k+1} - \log t_{k} = \text{const}$ between successive times $t_{k+1}$ and $t_k$. The time can then be calculated as $t_k = e^{k\, \Delta\!\log t}$. Moreover, we choose a logarithmically spaced momentum grid for the distribution function $\tilde{f}(t,\tilde{p})$, its derivative and the collision integral, in order to resolve the distribution function at very low momenta. The grid involves $N_p$ momenta between $\Lambda_\mathrm{IR}$ and $\Lambda_\mathrm{UV}$ such that the ratio between successive momenta is constant $\tilde{p}_{k+1}/\tilde{p}_k = \text{const}$. We employed $\Delta\!\log t = 0.1$, $\Lambda_\mathrm{IR} = 0.017$, $\Lambda_\mathrm{UV} = 17$ - $52$ and $N_p = 100$ for the plots of this section. 

For the computation of the collision integral and for the interpolation of the distribution function, we use methods\footnote{From Ref.~\cite{GSL:2009}, we frequently use the integration method \textit{gsl\_integration\_cquad}, which is particularly suitable for singular integrands. Such occur, for instance, in the real part of the ``one-loop'' retarded self-energy $\PiR$. The employed interpolation type is \textit{gsl\_interp\_akima}.} 
from the GNU Scientific Library~\cite{GSL:2009}. Interpolation is required since the integration methods need continuous functions for the integrand, and it is performed based on the sampling points $(\tilde{p}_k, \tilde{f}_{n,k} \equiv \tilde{f}(t_n,\tilde{p}_k))$ at time $t_n$. For momenta outside the momentum grid $[\Lambda_\mathrm{IR},\Lambda_\mathrm{UV}]$, we set the distribution function to zero. Therefore, some terms in the functional $\FFunc[\tilde{f}]$ in (\ref{eq_functional_F}) become zero when one of the momenta is outside of the momentum interval, and the collision integral loses its gain-minus-loss structure. To prevent this, we additionally set the whole functional $\FFunc[\tilde{f}]$ to zero in such cases to reduce deviations from particle number and energy density conservation. This leads to simplifications of integration boundaries within the collision integral~\eqref{eq_collision_integral_massaged}, which are further discussed in App.~\ref{sec_integration_boundaries}. 

In the numerical algorithm, we first initialize the distribution function $\tilde{f}_0$ and its momentum derivative $\tilde{f}'_0$ at the grid points $\tilde{p}_k$ at initial time. The time step $t_n \rightarrow t_{n+1}$ follows the explicit Euler method
\begin{align}
 \label{NUM_descrete_solver}
 \tilde{f}_{n+1,k} = \tilde{f}_{n,k}- \Delta\!\log t \left[ \alpha \tilde{f}_{n,k} + \beta \tilde{p}_k \tilde{f}'_{n,k} - C[\tilde{f}_n] \right].
\end{align}
Since the evaluation of the collision integral at a single momentum point $p_k$ neither depends on nor affects the evaluation at other momenta,
we parallelize the part of our solver, where the collision integral is computed for each momentum on the grid.

The accuracy of our solution algorithm is mainly limited by the interpolation of the derivative of the distribution function $\tilde{f}'$. For a typical functional form as in Eq.~\eqref{eq_fit_distribution_lattice}, the relative accuracy for our discretization was up to $10^{-2}$ as compared to the analytical expression of the derivative. Although increasing $N_p$ may improve the resolution, the computational costs will grow and we thus found a compromise that still provided sufficiently accurate results.

%%
%%%%%%%%%%%%%%%%%%%%%%%%%%%%%%%%%%%%%%%%%%%%%%%%%%%%%%%%%%%%%%%%%%%%%%%%%%%%%%%%%%%%%%%%%%%%%%%%%%%%
%%%%%%%%%%%%%%%%%%%%%%%%%%%%%%%%%%%%%%%%%%%%%%%%%%%%%%%%%%%%%%%%%%%%%%%%%%%%%%%%%%%%%%%%%%%%%%%%%%%%
%%
%%

%%%%%%%%%%%%%%%%%%%%%%%%%%%%%%%%%%%%%%%%%%%%%%%%%%%%%%%%%%%%%%%%%%%%%%%%%%%%%%%%%%%%%%%%%%%%%%%%%%%%
\begin{figure}[t]
	\centering
	\includegraphics[scale=0.23]{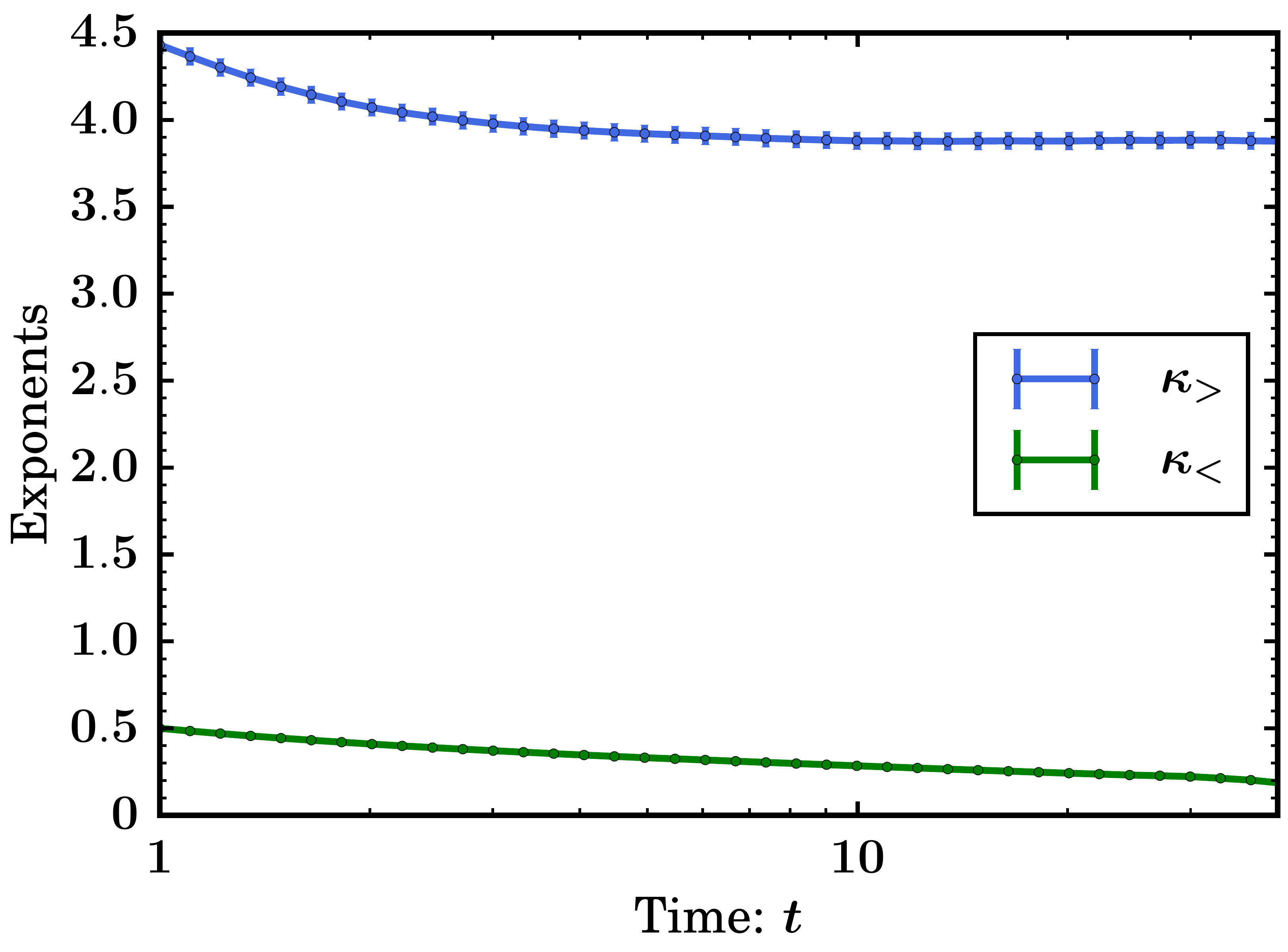}
	\caption{The exponents of approximate power laws $\tilde{p}^{-\kappa_i}$ at low ($\kappa_<$) and high ($\kappa_>$) momenta of the rescaled distribution $\tilde{f}$ as functions of time. The distribution has been initialized as in (\ref{eq_fit_distribution_lattice}) with $A = 27000$, $B=1.135$, $\kappa_< = 0.5$ and $\kappa_> = 4.5$. }
	\label{fig_KG_KS_evol}
\end{figure}
%%%%%%%%%%%%%%%%%%%%%%%%%%%%%%%%%%%%%%%%%%%%%%%%%%%%%%%%%%%%%%%%%%%%%%%%%%%%%%%%%%%%%%%%%%%%%%%%%%%%

\subsection{Details on the computation of the scaling function}
\label{sec_numerics_of_LNKT}

In Fig.~\ref{fig_alpha15_beta05_final_evol} we show the relaxation dynamics of the rescaled distribution $\tilde{f}$ computed by the algorithm introduced above. We start close to the stationary form by choosing $\tilde{f}_0$ as in Eq.~(\ref{eq_fit_distribution_lattice}) with $\kappa_< = 0$ and $\kappa_> = 3.9$. One observes that $\tilde{f}$ quickly approaches a stationary form, which can be understood as the scaling function $\fs$. Curves at times $t \geq 1.6$ are already almost time-independent. Therefore $\fs$ shown in Fig.~\ref{fig_alpha15_beta05_sol} is computed as the average over these curves, while the error bars are estimated by the standard deviation in this procedure. 

The functional form of $\fs$ is barely distinguishable from its starting form (\ref{eq_fit_distribution_lattice}) in Fig.~\ref{fig_alpha15_beta05_final_evol}, however, small deviations around the scale $\Ks$ exist. At low momenta $\tilde{p} \llSim \Ks$ and at high momenta $\tilde{p} \ggSim \Ks$, the scaling function follows power laws $\tilde{p}^{-\kappa_<}$ and $\tilde{p}^{-\kappa_>}$. We have measured the spectral exponents $\kappa_i$ by employing power law fits to the respective regions in the scaling function,
\begin{align}
 \label{eq_kappa_values}
 \kappa_< &= 0 \pm 0.01\; \textrm{(sys)} \nonumber \\
 \kappa_> &= 3.95 \pm 0.05\; \textrm{(sys)}\,.
\end{align}
Statistical errors are much smaller than systematic errors, which were estimated to contain possible infrared and ultraviolet cutoff artifacts.\footnote{We note that at low and high momenta close to the cutoffs, the scaling function starts to show deviations from power law or constant behavior, which is the main source of error in the power law fits. Increasing the momentum grid to lower and larger momenta reduces the effects of these numerical artifacts, but comes at the price of increased numerical costs.}  

To check the stability of these values, we start with slightly larger exponents $\kappa_< = 0.5$ and $\kappa_> = 4.5$ for the initial distribution $\tilde{f}_0$ with the functional form (\ref{eq_fit_distribution_lattice}). The time evolution of the exponents is shown in Fig.~\ref{fig_KG_KS_evol}, where we use the error estimates of (\ref{eq_kappa_values}). Indeed, one observes that the exponents approach the values (\ref{eq_kappa_values}) of the scaling function.

%%%%%%%%%%%%%%%%%%%%%%%%%%%%%%%%%%%%%%%%%%%%%%%%%%%%%%%%%%%%%%%%%%%%%%%%%%%%%%%%%%%%%%%%%%%%%%%%%%%%
%%%%%%%%%%%%		SECTION: CONCLUSION
%%%%%%%%%%%%%%%%%%%%%%%%%%%%%%%%%%%%%%%%%%%%%%%%%%%%%%%%%%%%%%%%%%%%%%%%%%%%%%%%%%%%%%%%%%%%%%%%%%%%
\section{Conclusion}
\label{sec_conclusion}

In this work, we have shown that the large-$N$ kinetic theory at NLO can be applied to highly occupied scalar quantum field theory, which cannot be described by a perturbative kinetic framework. On the other hand, for sufficiently low occupancies or at large momenta it effectively reduces to a perturbative kinetic theory at large $N$. Hence, the large-$N$ kinetic theory extends perturbative descriptions and constitutes a versatile tool to study the dynamics of systems out of equilibrium in terms of quasiparticles.

An essential ingredient for the quasiparticle picture, for free movement between collisions and for the proper inclusion of quantum interference effects even at high occupancies is the vertex resummation that appears at NLO in the large-$N$ expansion~\cite{Berges:2001fi,Aarts:2002dj,Scheppach:2009wu,Berges:2010ez,Orioli:2015dxa}. We analyzed the structure of the effective vertex in detail, which enabled us to show that the mean free path $L$ of quasiparticles at typical momentum modes is larger than their de Broglie wave length. Moreover, the spectral function can be approximated by a quasiparticle form at NLO of the large-$N$ theory since a peak width is suppressed by $1/N$. These arguments lead to a well-defined dispersion relation, and thus a consistent quasiparticle picture. 

In a second step, we applied the large-$N$ kinetic theory to the highly occupied region of scalar systems at low momenta characterized by a universal self-similar evolution~\cite{Orioli:2015dxa,Moore:2015adu}. Surpassing former analytical estimates for the scaling exponents $\alpha$ and $\beta$ of the self-similar evolution \cite{Orioli:2015dxa}, we solved the effective kinetic equation numerically for the first time. The scaling function obtained, $\fs(|\mathbf{p}|)$, agrees well with former lattice simulation results, which is a striking confirmation of the applicability of the large-$N$ kinetic theory to highly occupied systems. It reveals a power law behavior $\sim |\mathbf{p}|^{-4}$ at momenta higher than the typical momentum $\Ks$ that dominates particle number density, and becomes constant at low momenta. 

No explicit assumption on the coupling strength has entered the derivation of the large-$N$ kinetic theory. Therefore, in principle, one could use the large-$N$ kinetic theory also at moderate couplings for finite $N$. Indeed, it was shown using 2PI equations to NLO in a $1/N$ expansion \cite{Berges:2001fi,Aarts:2002dj,Scheppach:2009wu} that the observed self-similar regime at low momenta survives to moderate values $\lambda = 1$ \cite{Berges:2016nru}. However, it was argued within the 2PI framework that for a large coupling $\lambda = 10$, inelastic processes may play an important role at all times in the evolution \cite{Tsutsui:2017uzd}. The large-$N$ kinetic theory in its present form at NLO, however, lacks such processes. 

To be able to simulate a complete thermalization process within large-$N$ kinetic theory, starting far from equilibrium and the evolution towards a nonthermal fixed point including the subsequent final thermalization dynamics, one would have to go beyond this order to capture inelastic processes. While going to NNLO is challenging, the description may be partially simplified at late times relevant for the final approach to thermal equilibrium since the typical occupancies become smaller such that standard perturbative approximations become available again at least for small enough couplings. 

The large-$N$ kinetic theory is an example for a kinetic theory applicable to highly occupied systems, for which conventional kinetic approaches fail. For non-Abelian plasmas, multiple studies indicate strong fields and nontrivial dynamics at low-momentum modes \cite{Berges:2015ixa,Berges:2013eia,Berges:2013fga,Tanji:2017suk,Mace:2016svc,Berges:2017igc}. An effective description thereof could be an important extension of kinetic approaches \cite{Baier:2000sb,Blaizot:2001nr,Arnold:2002zm,Kurkela:2014tea,Kurkela:2015qoa} to the evolution of weakly coupled non-Abelian plasmas and the thermalization process in ultrarelativistic heavy-ion collisions at high energies.

%%%%%%%%%%%%%%%%%%%%%%%%%%%%%%%%%%%%%%%%%%%%%%%%%%%%%%%%%%%%%%%%%%%%%%%%%%%%%%%%%%%%%%%%%%%%%%%%%%%%
%%%%%%%%%%%%		SECTION: ACKNOWLEDGEMENTS
%%%%%%%%%%%%%%%%%%%%%%%%%%%%%%%%%%%%%%%%%%%%%%%%%%%%%%%%%%%%%%%%%%%%%%%%%%%%%%%%%%%%%%%%%%%%%%%%%%%%

\begin{acknowledgments}
We thank J.~P.~Blaizot, I.~Chantesana, T.~Gasenzer, A.~Kurkela, T.~Lappi, A.~Pi\~neiro Orioli, S.~Schlichting and R.~Venugopalan for useful discussions and collaborations on related work. K.B.~gratefully acknowledges support by the European Research Council under grant No.~ERC-2015-COG-681707. This work is part of and supported by the DFG Collaborative Research Centre ``SFB 1225 (ISOQUANT)".
\end{acknowledgments}

%%%%%%%%%%%%%%%%%%%%%%%%%%%%%%%%%%%%%%%%%%%%%%%%%%%%%%%%%%%%%%%%%%%%%%%%%%%%%%%%%%%%%%%%%%%%%%%%%%%%
%%%%%%%%%%%%		APPENDIX
%%%%%%%%%%%%%%%%%%%%%%%%%%%%%%%%%%%%%%%%%%%%%%%%%%%%%%%%%%%%%%%%%%%%%%%%%%%%%%%%%%%%%%%%%%%%%%%%%%%%

\appendix

%%%%%%%%%%%%%%%%%%%%%%%%%%%%%%%%%%%%%%%%%%%%%%%%%%%%%%%%%%%%%%%%%%%%%%%%%%%%%%%%%%%%%%%%%%%%%%%%%%%%
%%%%%%%%%%%%		SECTION: ANALYTICAL PREPARATION OF COLLISION INTEGRAL AND SELF-ENERGY
%%%%%%%%%%%%%%%%%%%%%%%%%%%%%%%%%%%%%%%%%%%%%%%%%%%%%%%%%%%%%%%%%%%%%%%%%%%%%%%%%%%%%%%%%%%%%%%%%%%%

\section{Towards solving the large-$N$ kinetic equation}
\label{sec_preparation}

In this Appendix, we solve some of the integrals appearing in the collision integral of the non-relativistic large-$N$ kinetic theory (\ref{eq_collision_integral_raw}) analytically for $d = 3$ spatial dimensions.

\subsection{Collision integral}
\label{sec_preparation_coll_int}

We start by averaging the Boltzmann equation (\ref{eq_effective_Boltzmann_equation}) over the solid angle of $\mathbf{p}$. Because of the isotropy of the distribution function, the left hand side does not change while the right hand side becomes
\begin{align}
C[f](t,p) \,&= \qquad \int\frac{\mathrm{d} \Omega_\mathbf{p}}{4\pi} \, C[f](t,\mathbf{p}) \nonumber\\
&= \frac{\pi}{(2\pi)^{10}} \int_0^{\infty} \mathrm{d} l \, l^2 \, \mathrm{d} q \, q^2 \, \mathrm{d} r \, r^2 \int \mathrm{d} \CKernel(p,l,q,r) \nonumber\\
& \qquad \qquad \times \delta\left( \omega_p + \omega_l - \omega_q - \omega_r \right) \nonumber\\
& \qquad \qquad \times g_\mathrm{eff}^2[f](t,\omega_p - \omega_q,\mathbf{\pMq}) \nonumber\\
& \qquad \qquad \times \FFunc[f](t,p,l,q,r). 
\label{LNKT_collision_integral_raw}
\end{align}
We have introduced the momentum difference $\mathbf{\pMq} \equiv \mathbf{p-q}$ for the effective coupling $g_\mathrm{eff}^2$. In App.~\ref{sec_preparation_self_energy} we will show that it only depends on the magnitude $\pMq = \sqrt{p^2 + q^2 - 2pq \cos(\theta_{\mathbf{p,q}})}$, and hence on the magnitudes of the momenta $p = |\mathbf{p}|$ and $q = |\mathbf{q}|$ and on the angle $\theta_\mathbf{p,q}$ between them. The three-dimensional momentum integrals in (\ref{LNKT_collision_integral_raw}) have been split into radial and solid angle parts according to $\int \mathrm d ^3 q = \int_{0}^{\infty} \mathrm d q \, q^2 \int \mathrm d \Omega_\mathbf{q}$ with $\int \mathrm d \Omega_\mathbf{q} = \int_{0}^{2\pi} \mathrm d \varphi_\mathbf{q} \int_{-1}^{1} \mathrm d \cos\left(\theta_{\mathbf{q}}\right)$. All angular integrals have been included in
\begin{align}
\int \mathrm{d} \CKernel(p,l,q,r) \equiv &\int \mathrm{d}\Omega_{\mathbf{p}} \, \mathrm{d}\Omega_{\mathbf{l}} \, \mathrm{d}\Omega_{\mathbf{q}}\, \mathrm{d}\Omega_{\mathbf{r}} \nonumber\\
& \times (2\pi)^3\delta^{(3)}(\mathbf{p+l-q-r})
\label{LNKT_function_K}
.
\end{align}
Except for $\theta_\mathbf{p,q}$, there is no angular dependence in the residual terms of the collision integral. In the following, the respective integrals are performed analytically. 

%%%%%%%%%%%%%%%%%%%%%%%%%%%%%%%%%%%%%%%
\begin{figure}[t]
	\centering
	\includegraphics[scale=0.34]{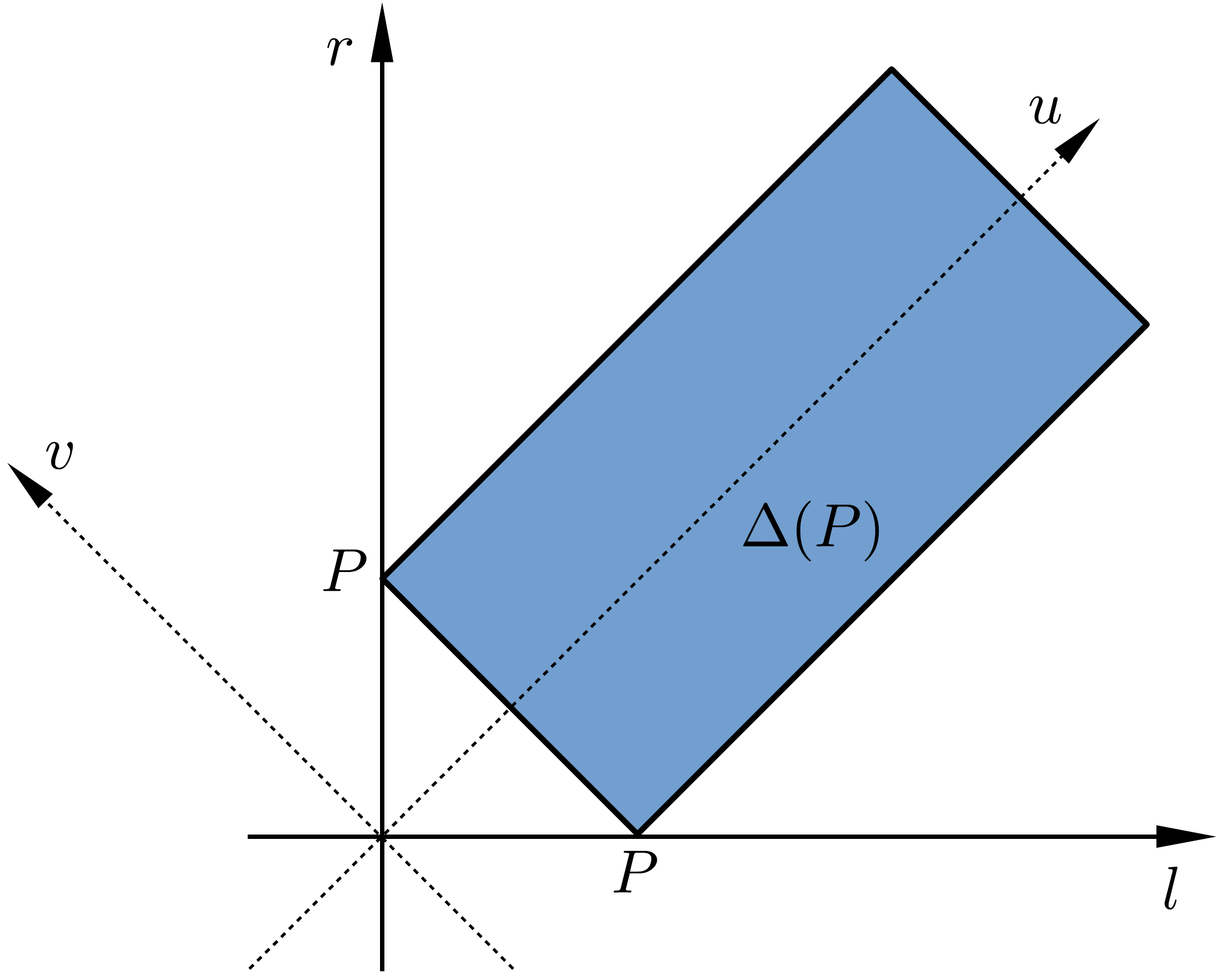}
	\caption{The filled region $\Delta(\pMq)$ represents the area of integration in~\eqref{LNKT_integration_sign_functions}.
		The figure is taken with adaption from Ref.~\cite{Orioli:2015dxa}.}
	\label{fig_int_region_for_sign_functions}
\end{figure}
%%%%%%%%%%%%%%%%%%%%%%%%%%%%%%%%%%%%%%%

We start by using the integral representation of the delta function
\begin{equation}
(2\pi)^3\delta^{(3)}(\mathbf{p+l-q-r}) = \int \mathrm{d}^3 x \, e^{i\mathbf{(p+l-q-r)x}}.
\label{LNKT_integral_representation_delta_function}
\end{equation}
Exploiting $\int \mathrm{d} \Omega_{\mathbf{p}} \mathrm{d} \Omega_{\mathbf{q}} = \int \mathrm{d} \Omega_{\mathbf{q}} \mathrm{d} \Omega_{\mathbf{p,q}}$, where $\int \mathrm{d} \Omega_{\mathbf{p,q}}$ denotes the angular integration of $\mathbf{p}$ around the axis in $\mathbf{q}$ direction, and performing integrations over the angles except for $\theta_{\mathbf{p,q}}$, we arrive at
\begin{align}
\int \mathrm{d} \CKernel(p,l,q,r) = & \,\, 2^4 (2\pi)^5 \int_{-1}^{1} \mathrm{d}\cos(\theta_\mathbf{p,q}) \int_0^{\infty} \mathrm{d} x \, x^2 \nonumber\\
& \times \frac{\sin(\pMq x)}{\pMq x} \, \frac{\sin(lx)}{lx} \, \frac{\sin(rx)}{rx},
\label{APP_ANG_K1}
\end{align}
where we employed $\int_{-1}^{1} \mathrm{d} y \, e^{\pm ixy} = 2 \sin(x) / x$. We have used that the collision integral does not depend on the polar angle $\theta_{\mathbf{q}}$ of $\mathbf{q}$, even after integrating over $\theta_\mathbf{p,q}$, such that the integration $\int_{-1}^{1} \mathrm{d}\cos(\theta_{\mathbf{q}}) = 2$ can be performed in the end, which leads to Eq.~\eqref{APP_ANG_K1}. With
\begin{align}
&\; \sin(a) \sin(b) \sin(c) \nonumber\\
= &\,\frac{1}{4} \left( -\sin(a-b-c) + \sin(a+b-c)\right. \nonumber\\
&\quad+ \left. \sin(a-b+c) - \sin(a+b+c) \right)
\label{eq_three_sinuses}
\end{align}
and
\begin{equation}
\int_0^{\infty} \mathrm{d} x \, \frac{\sin(ax)}{x} = \frac{\pi}{2} \, \mathrm{sgn}(a)
\label{eq_sinc_integral}
\end{equation}
we obtain
\begin{align}
\int \mathrm{d} \CKernel&(p,l,q,r) = \frac{(2\pi)^6}{l r} \int_{-1}^{1} \frac{\mathrm{d}\cos(\theta_{\mathbf{p,q}})}{\pMq} \nonumber\\
\times \Big( &\mathrm{sgn}(\pMq +l-r) + \mathrm{sgn}(\pMq - l + r) \nonumber \\
- &\mathrm{sgn}(\pMq -l-r) - \mathrm{sgn}(\pMq + l + r ) \Big)
\label{LNKT_function_K_2}
.
\end{align}
Taking the integrals over $l$ and $r$ of the collision integral~\eqref{LNKT_collision_integral_raw} into account, the sign functions in expression~\eqref{LNKT_function_K_2} can be conveniently evaluated via
\begin{align}
	&\int_{0}^{\infty} \mathrm{d} l \int_{0}^{\infty} \mathrm{d} r \Big( \mathrm{sgn}(\pMq +l-r) + \mathrm{sgn}(\pMq - l + r) \nonumber \\
	& \qquad \qquad \quad ~~ -\mathrm{sgn}(\pMq-l-r) - \mathrm{sgn}(\pMq + l + r) \Big) \nonumber\\
	= \,&2 \left( \int_{\pMq}^{\infty} \mathrm{d} r \int_{0}^{\pMq+r} \mathrm{d} l + \int_{0}^{\pMq} \mathrm{d} r \int_{\pMq-r}^{\pMq+r} \mathrm{d} l \right. \nonumber \\
	& ~\; \left. - \int_{0}^{\infty} \mathrm{d} l \int_{\pMq+l}^{\infty} \mathrm{d} r \right) \nonumber \\
	\equiv \,&2 \int_{\Delta(\pMq)} \mathrm{d} l \, \mathrm{d} r ~=~ \int_{\pMq}^{\infty} \mathrm{d} u \int_{-\pMq}^{\pMq} \mathrm{d} v
	\label{LNKT_integration_sign_functions}
	.
\end{align}
A change of variables from $l$ and $r$ to $u=r+l$ and $v=r-l$ has been performed in the last step of Eq.~\eqref{LNKT_integration_sign_functions} absorbing a factor of 2.
Fig.~\ref{fig_int_region_for_sign_functions} visualizes the region of integration $\Delta(\pMq)$.
A second transformation from $\mathrm d \cos(\theta_\mathbf{p,q})$ to $\mathrm d \pMq$ with $\mathrm{d} \cos(\theta_\mathbf{p,q}) = - (\pMq / pq ) \, \mathrm{d} \pMq$ leads to
\begin{align}
C[f]&(t,p) = \frac{1}{64 \pi^3 p} \int_{0}^{\infty} \mathrm{d} q \int^{p+q}_{\left|p-q\right|} \mathrm{d} \pMq 
\int_{\pMq}^{\infty} \mathrm{d} u \int_{-\pMq}^{\pMq} \mathrm{d} v
\nonumber\\
& \times
q \left(u^2 - v^2\right) 
\delta\left(\omega_p + \omega_{(u-v)/2} - \omega_q - \omega_{(u+v)/2} \right)
\nonumber\\
& \times
g_{\mathrm{eff}}^2[f](t,\omega_p - \omega_q,\pMq) 
\nonumber\\
& \times
\FFunc[f]\left(t,p,\frac{u-v}{2},q,\frac{u+v}{2}\right).
\label{LNKT_solid_angle_coll_int}
\end{align}

The next step is the evaluation of the energy-conserving delta function. With the quadratic dispersion relation $\omega_p = p^2/2m$, the delta function in the collision integral~\eqref{LNKT_solid_angle_coll_int} reads
\begin{align}
&\delta \left( \frac{1}{2m} \left( p^2 + \frac{1}{4}\left( u - v \right)^2 - q^2
- \frac{1}{4} \left( u + v \right)^2 \right) \right)
\nonumber\\
&= 2m \, \delta \left( p^2 - q^2 - uv \right).
\end{align}
We use the $v$-integral to evaluate the delta function according to
\begin{align}
2m &\int_{-\pMq}^{\pMq} \mathrm{d} v \, R(v) \, \delta \left( p^2 - q^2 - uv \right)
\nonumber\\
&= \frac{2m}{u} \, R\left( \frac{p^2 - q^2}{u} \right) \Theta \left( \pMq - \frac{ \left| p^2 - q^2 \right| }{u} \right)
,
\label{LNKT_evaluating_delta_function}
\end{align}
where $R$ represents every part of the collision integral depending on $v$. The step function in (\ref{LNKT_evaluating_delta_function}) can be used to change the integration boundaries of $u$ to $\int_{\mathrm{max}\left(\pMq, \left|p^2-q^2\right| / \pMq \right)}^{\infty} \mathrm{d} u$. The resulting form of the collision integral is given by Eq.~(\ref{eq_collision_integral_massaged}).

\subsection{Retarded self-energy}
\label{sec_preparation_self_energy}

To also simplify the computation of the effective coupling $g_{\mathrm{eff}}^2$ of (\ref{eq_geff}), we perform the angular integrations of the ``one-loop'' retarded self-energy $\PiR$ in (\ref{eq_1loop_ret_self_energy_raw}) analytically. We first change the integration variable to $\mathbf{\pMq - k} \mapsto \mathbf{k}$. Introducing $y \equiv \cos(\theta_\mathbf{\pMq,k})$, where $\theta_\mathbf{\pMq,k}$ is the polar angle between the momenta $\mathbf{\pMq}$ and $\mathbf{k}$, one arrives at 
\begin{align}
\label{eq_PiR_with_y_variable}
\PiR(t,\omega,\pMq) = \lim\limits_{\epsilon \to 0^{+}} & \frac{g}{(2\pi)^2} \int_{0}^{\infty} \mathrm d k \, k^2 f(t,k) \int_{-1}^{1}\mathrm d y \nonumber\\
\times \Bigg[ & \frac{1}{(\pMq^2-2 \pMq ky)/2m - \omega - i\epsilon} \nonumber\\
+ \,& \frac{1}{(\pMq^2-2 \pMq ky)/2m + \omega + i\epsilon} \Bigg].
\end{align}
Note that due to the isotropy of $f(t,k)$, $\PiR$ only depends on the absolute value of the momentum $\pMq = |\mathbf{\pMq}|$. Moreover, it vanishes for $\pMq = 0$ since the integrand becomes identically zero. In addition, one observes that changing the frequency from $\omega$ to $-\omega$ corresponds to complex conjugation of the whole expression. As a consequence, the ``one-loop'' retarded self-energy $\PiR$ is real for $\omega = 0$.

Rewriting (\ref{eq_PiR_with_y_variable}) as
\begin{align}
\PiR & (t,\omega,\pMq)
=
\lim\limits_{\tilde \epsilon \to 0^{+}} \frac{-mg}{(2\pi)^2 \pMq} \int_0^\infty \mathrm d k \, k \, f(t,k)
\int_{-1}^{1}\mathrm{d} y
\nonumber\\
& \times
\left[ \frac{1}{y - \frac{\pMq^2 - 2m\omega}{2 \pMq k} + i\tilde\epsilon} 
+ \frac{1}{y - \frac{\pMq^2 + 2m\omega}{2 \pMq k} - i\tilde\epsilon} \right]
,
\label{LNKT_before_PV_calculation}
\end{align}
where we have substituted $\tilde{\epsilon} \equiv m\epsilon/\pMq k$, enables us to use the principal value integral
\begin{align}
\lim_{\epsilon \to 0^+} & \int_{-1}^{1} \frac{1}{x+D \pm i\epsilon} \, \mathrm d x 
\nonumber\\
&= \mathrm{PV} \int_{-1}^{1} \frac{1}{x+D} \, \mathrm d x \, \mp i \pi \int_{-1}^{1} \delta(x+D) \, \mathrm d x
\nonumber\\
&= \log \left| \frac{1 + D}{-1+D} \right| \quad \quad \mp i \pi \, \Theta( 1 - |D|)
\label{LNKT_PV_integral}
\end{align}
for each fraction in Eq.~\eqref{LNKT_before_PV_calculation}.
This leads to
\begin{equation}
\PiR(t,\omega,\pMq)
= \frac{-mg}{(2\pi)^2 \pMq} \int_{0}^{\infty} \mathrm d k \, k \, f(t,k) \, \PiKernel(\omega,\pMq,k)
\end{equation}
with the kernel
\begin{align}
&\PiKernel(\omega,\pMq,k)
=
\log \left| \frac{\left(\pMq^2 - 2 \pMq k\right)^2 - 4m^2\omega^2}{\left(\pMq^2 + 2 \pMq k\right)^2 - 4m^2\omega^2} \right|
\nonumber\\
&- i\pi
\left[ \Theta\left( 1 - \frac{\left| \pMq^2 - 2m\omega \right|}{2 \pMq k} \right) 
- \Theta \left( 1 - \frac{\left| \pMq^2 + 2m\omega \right|}{2 \pMq k} \right) \right]
\label{eq_integrand_self_energy}
.
\end{align}

For a numerical treatment it is useful to know the singular points of the real part of the integrand $\mathrm{Re} \, \PiKernel(\omega,\pMq,k)$ and the region where the imaginary part $\mathrm{Im} \, \PiKernel(\omega,\pMq,k)$ does not vanish. The singularities of the real part are given by 
\begin{align}
 k_{\textrm{sing}} = \pm\frac{\pMq}{2} \pm \frac{m|\omega|}{\pMq}\,,
\end{align}
where the case $\pMq = 0$ is excluded since the whole integrand is zero then. The real part can be rewritten as
\begin{align}
\mathrm{Re} \, \PiKernel(\omega,\pMq,k) = \log\left| \frac{ \left(k - \frac{\pMq}{2}\right)^2 - \frac{m^2 \omega^2}{\pMq^2} } { \left(k + \frac{\pMq}{2}\right)^2 - \frac{m^2 \omega^2}{\pMq^2} } \right|.
\end{align}
For the imaginary part $\mathrm{Im} \, \PiKernel(\omega,\pMq,k)$ of the integrand~\eqref{eq_integrand_self_energy}, we will assume $\omega > 0$ since the case of a negative frequency is related to the positive frequency case by complex conjugation as noted above. Then, $\mathrm{Im} \, \PiKernel(\omega,\pMq,k)$ is zero until the integration variable $k$ has increased sufficiently to fulfill the condition $k \geq |\pMq^2 - 2m\omega|/2\pMq$ of the first Heaviside step function but is still too small to fulfill the condition of the second step function. After exceeding $|\pMq^2 + 2m\omega|/2\pMq$, which makes the second step function one as well, the whole expression vanishes again. Altogether, we find
\begin{equation}
-\frac{1}{\pi} \mathrm{Im} \, \PiKernel(\omega,\pMq,k) 
= 
\begin{cases}
1 &\text{if } \frac{|\pMq^2 - 2m\omega|}{2\pMq} \leq k \leq \frac{|\pMq^2 + 2m\omega|}{2\pMq} \\
0 &\text{else } 
\end{cases}
\!.
\end{equation}
Including also the case of negative frequency, we arrive at the final form of the ``one-loop'' retarded self-energy in Eq.~(\ref{eq_1loop_ret_self_energy_massaged}).

%%%%%%%%%%%%%%%%%%%%%%%%%%%%%%%%%%%%%%%%%%%%%%%%%%%%%%%%%%%%%%%%%%%%%%%%%%%%%%%%%%%%%%%%%%%%%%%%%%%%
%%%%%%%%%%%%		APPENDIX INTEGRATION BOUNDARIES
%%%%%%%%%%%%%%%%%%%%%%%%%%%%%%%%%%%%%%%%%%%%%%%%%%%%%%%%%%%%%%%%%%%%%%%%%%%%%%%%%%%%%%%%%%%%%%%%%%%%
\section{Integration boundaries}
\label{sec_integration_boundaries}

As explained in Sec.~\ref{sec_numerical_setup}, the collision integral $C[f](t,p)$ as well as the distribution function $f(t,p)$ are discretized on a grid between the momenta $\Lambda_{\mathrm{IR}}$ and $\Lambda_{\mathrm{UV}}$ in our numerical approach. Outside this domain, we set $f(t,p)=0$. Making use of this, the integration boundaries of the collision integral~\eqref{eq_collision_integral_massaged} can be further constrained. 

Integrals within the real and imaginary part of the ``one-loop'' retarded self-energy~\eqref{eq_1loop_ret_self_energy_massaged} are simplified to
\begin{alignat}{2}
	&\mathrm{Re} \, \PiR: \int_{0}^{\infty} \mathrm d k &&\to
	\int_{\Lambda_\mathrm{IR}}^{\Lambda_\mathrm{UV}} \mathrm d k, 
	\label{NUM_int_bound_real}
	\\
	&\mathrm{Im} \, \PiR: \int_{\frac{|\pMq^2 - 2m\omega|}{2\pMq}}^{\frac{|\pMq^2 + 2m\omega|}{2\pMq}} \mathrm d k &&\to
	\int_{\mathrm{max} \left( \frac{|\pMq^2 - 2m\omega|}{2\pMq}, \, \Lambda_\mathrm{IR} \right)}
	^{\mathrm{min} \left( \frac{|\pMq^2 + 2m\omega|}{2\pMq}, \, \Lambda_\mathrm{UV} \right)} \mathrm d k.
\end{alignat}
The gain-minus-loss part
\begin{align}
\FFunc[f] \left(t, p, \frac{1}{2}\left( u - \frac{p^2 - q^2}{u} \right), q, \frac{1}{2}\left( u + \frac{p^2 - q^2}{u} \right) \right)
\end{align}
of the integrand of the collision integral yields non-vanishing contributions if
\begin{alignat}{2}
\Lambda_\mathrm{IR} & \leq p && \leq \Lambda_\mathrm{UV}, \quad \nonumber\\
\Lambda_\mathrm{IR} & \leq q && \leq \Lambda_\mathrm{UV}, \quad \nonumber\\
\Lambda_\mathrm{IR} & \leq \frac{1}{2}\left( u \pm \frac{p^2 - q^2}{u} \right) && \leq \Lambda_\mathrm{UV}
.
\label{eq_condition_nonvanishing_coll_int}
\end{alignat}
To preserve the gain-minus-loss symmetry, we set the whole $\FFunc$ to zero if one of the conditions (\ref{eq_condition_nonvanishing_coll_int}) is not fulfilled, as explained in Sec.~\ref{sec_numerical_setup}. The first constraint is automatically fulfilled since the momentum $p$ is externally set to the correct momentum range. The $q$-integration is changed to
\begin{equation}
\int_0^\infty \mathrm d q \to \int_{\Lambda_\mathrm{IR}}^{\Lambda_\mathrm{UV}} \mathrm d q\,.
\end{equation}
The relation
\begin{equation}
\Lambda_\mathrm{IR} \leq \frac{1}{2}\left( u \pm \frac{\pSqrMqSqr}{u} \right) \leq \Lambda_\mathrm{UV}
\label{NUM_rel_for_u_boundaries}
\end{equation}
with $\pSqrMqSqr \equiv \left| p^2 - q^2 \right|$ is chosen to be fulfilled for plus and minus sign simultaneously. It can be solved for $u$ to obtain new integration boundaries according to the following procedure:

In a first step, we consider the left inequality of~\eqref{NUM_rel_for_u_boundaries}, i.e.\ $(u \pm \pSqrMqSqr/u) / 2 \geq \Lambda_\mathrm{IR}$. Using that $u>0$, the inequality can be transformed to
\begin{align}
 \label{NUM_rel_for_u_IR_mp}
 \left( u - \Lambda_\mathrm{IR} \right)^2 \geq \Lambda_\mathrm{IR}^2 \mp \pSqrMqSqr\,,
\end{align}
where the two cases of minus and plus sign on the right-hand side are distinguished.

	For the case of the minus sign, no restriction for $u$ is obtained if $\Lambda_\mathrm{IR}^2 \leq \pSqrMqSqr$. Otherwise, one obtains
	\begin{align}
		u \leq \Lambda_\mathrm{IR} - \sqrt{\Lambda_\mathrm{IR}^2 - \pSqrMqSqr} \qquad & \mbox{if } u < \Lambda_\mathrm{IR}
		\label{NUM_rel_for_u_1}
		,
		\\
		u \geq \Lambda_\mathrm{IR} + \sqrt{\Lambda_\mathrm{IR}^2 - \pSqrMqSqr} \qquad & \mbox{if } u > \Lambda_\mathrm{IR}
		\label{NUM_rel_for_u_2}	
		.
	\end{align}
	In case of the plus sign on the right-hand side of (\ref{NUM_rel_for_u_IR_mp}), one finds the constraints
	\begin{align}
		\qquad \, \, u \geq \Lambda_\mathrm{IR} + \sqrt{\Lambda_\mathrm{IR}^2 + \pSqrMqSqr} \qquad & \mbox{if } u > \Lambda_\mathrm{IR}
		\label{NUM_rel_for_u_3}
		,
		\\
		u \leq \Lambda_\mathrm{IR} - \sqrt{\Lambda_\mathrm{IR}^2 + \pSqrMqSqr} \qquad & \mbox{if } u < \Lambda_\mathrm{IR}
		\label{NUM_rel_for_u_not_needed}
		.
	\end{align}
	The latter condition~\eqref{NUM_rel_for_u_not_needed} excludes $u < \Lambda_\mathrm{IR}$ since $u$ has always to be positive. 

In a second step, we consider the inequality on the right hand side of (\ref{NUM_rel_for_u_boundaries}), which can be transformed to 
\begin{align}
 \left|u - \Lambda_\mathrm{UV}\right| \leq \sqrt{ \Lambda_\mathrm{UV}^2 \mp \pSqrMqSqr }\,.
\end{align}
This leads to the four conditions
\begin{alignat}{2}
	u &\geq \Lambda_\mathrm{UV} - \sqrt{\Lambda_\mathrm{UV}^2 - \pSqrMqSqr} \qquad &&\text{if } u < \Lambda_\mathrm{UV}
	\label{NUM_rel_for_u_4}
	,
	\\
	u &\leq \Lambda_\mathrm{UV} + \sqrt{\Lambda_\mathrm{UV}^2 - \pSqrMqSqr} \qquad &&\text{if } u > \Lambda_\mathrm{UV}
	\label{NUM_rel_for_u_5}
	,
	\\
	u &\leq \Lambda_\mathrm{UV} + \sqrt{\Lambda_\mathrm{UV}^2 + \pSqrMqSqr} \qquad &&\text{if } u > \Lambda_\mathrm{UV}
	\label{NUM_rel_for_u_6}
	,
	\\
	u &\geq \Lambda_\mathrm{UV} - \sqrt{\Lambda_\mathrm{UV}^2 + \pSqrMqSqr} \qquad &&\text{if } u < \Lambda_\mathrm{UV}
	\label{NUM_rel_for_u_not_needed_2}
	,
\end{alignat}
where the condition~\eqref{NUM_rel_for_u_not_needed_2} is no constraint, since $\Lambda_\mathrm{UV} - \sqrt{\Lambda_\mathrm{UV}^2 + \pSqrMqSqr} \leq 0$. 

All considered cases and the successive constraints are employed at the same time. This allows us to change the integration range for $u$ to
\begin{align}
&\int_{\mathrm{max}\left(\pMq, \, \frac{\pSqrMqSqr}{\pMq}\right)}^{\infty} \mathrm d u \nonumber\\
\to &\int_{\mathrm{max}\left(\pMq, \, \frac{\pSqrMqSqr}{\pMq}, \, \Lambda_\mathrm{IR} + \sqrt{\Lambda_\mathrm{IR}^2 + \pSqrMqSqr}, \,
	\Lambda_\mathrm{UV} - \sqrt{\Lambda_\mathrm{UV}^2 - \pSqrMqSqr}\right)}^{\Lambda_\mathrm{UV} + \sqrt{\Lambda_\mathrm{UV}^2 - \pSqrMqSqr}}
\mathrm d u
\label{NUM_changed_u_integration}
.
\end{align}

To have a nontrivial integration range for $u$, its integration boundaries should additionally satisfy
\begin{align}
 \label{NUM_upper_gtr_lower_u_integr}
 \Lambda_\mathrm{UV} + \sqrt{\Lambda_\mathrm{UV}^2 - \pSqrMqSqr} \geq \mathrm{max}\left(\pMq, \frac{\pSqrMqSqr}{\pMq}\right),
\end{align}
while the other two arguments of the maximum function in the lower integration boundary of $u$ are always smaller than the upper integration boundary. Eq.~(\ref{NUM_upper_gtr_lower_u_integr}) imposes constraints on the $\pMq$-integration. Thus, we change the boundaries of the $\pMq$-integration according to
\begin{align}
&\int_{|p - q|}^{p + q} \mathrm d \pMq \nonumber \\
\to &\int_{\mathrm{max} \left( |p - q|, \, | p^2 - q^2 |/\left(\Lambda_\mathrm{UV} + \sqrt{\Lambda_\mathrm{UV}^2 - | p^2 - q^2 |}\right) \right)}^{\mathrm{min} \left( p + q, \, \Lambda_\mathrm{UV} + \sqrt{\Lambda_\mathrm{UV}^2 - | p^2 - q^2 |}\right)} \mathrm d \pMq.
\end{align}
Furthermore, $u$- and $\pMq$-integrations are performed only if the lower integration boundaries are smaller than the upper ones. Otherwise, these integrals are set to zero.

%%%%%%%%%%%%%%%%%%%%%%%%%%%%%%%%%%%%%%%%%%%%%%%%%%%%%%%%%%%%%%%%%%%%%%%%%%%%%%%%%%%%%%%%%%%%%%%%%%%%
%%%%%%%%%%%%		BIBLIOGRAPHY
%%%%%%%%%%%%%%%%%%%%%%%%%%%%%%%%%%%%%%%%%%%%%%%%%%%%%%%%%%%%%%%%%%%%%%%%%%%%%%%%%%%%%%%%%%%%%%%%%%%%

\bibliography{Bibliography}

\end{document}